\documentclass[journal=jctcce,manuscript=article]{achemso}

\usepackage[version=3]{mhchem} %
\usepackage{graphicx}
\graphicspath{{Figures/}}
\usepackage{subcaption}
\usepackage{siunitx}
\DeclareSIUnit\Molar{M}

\usepackage{xcolor}

\usepackage{algorithm}
\usepackage[noend]{algpseudocode}
\usepackage{xpatch}
\algrenewcommand\alglinenumber[1]{{\sffamily\footnotesize#1}}
\makeatletter
\xpatchcmd{\algorithmic}{\itemsep\z@}{\itemsep=1ex plus1pt}{}{}
\makeatother
\author{Nicola Calonaci}
\affiliation[SISSA]{Scuola Internazionale Superiore di Studi Avanzati, SISSA, via Bonomea 265, Trieste 34136, Italy}
\alsoaffiliation[UNITS]{Department of Mathematics and Geosciences, University of Trieste, Trieste 34127, Italy}
\author{Mattia Bernetti}
\affiliation[SISSA]{Scuola Internazionale Superiore di Studi Avanzati, SISSA, via Bonomea 265, Trieste 34136, Italy}
\author{Alisha Jones}
\affiliation{Institute of Structural Biology, Helmoltz Zentrum Mu\"{u}nchen, Neuherberg 85764}
\alsoaffiliation{Bavarian NMR Center at Department of Chemistry, Technical University of Munich, Garching 85757, Germany}
\author{Michael Sattler}
\affiliation{Institute of Structural Biology, Helmoltz Zentrum Mu\"{u}nchen, Neuherberg 85764}
\alsoaffiliation{Center for Integrated Protein Science M\"{u}nchen and Bavarian NMR Center at Department of Chemistry, Technical University of Munich, Garching 85757, Germany}
\author{Giovanni Bussi}
\affiliation[SISSA]{Scuola Internazionale Superiore di Studi Avanzati, SISSA, via Bonomea 265, Trieste 34136, Italy}
\email{bussi@sissa.it}
\title[An \textsf{achemso} demo]
  {Molecular dynamics simulations with grand-canonical reweighting suggest cooperativity effects in RNA structure probing experiments}

\abbreviations{IR,NMR,UV}

\begin{document}
\begin{abstract}

Chemical probing experiments such as SHAPE are routinely used to probe RNA molecules.
In this work, we use atomistic molecular dynamics simulations to test the hypothesis that binding of RNA with SHAPE reagents
is affected by cooperative effects leading to an observed reactivity that is dependent on the reagent concentration.
We develop a general technique that enables the calculation of the affinity for arbitrary molecules as a function of their concentration in the grand-canonical ensemble.
Our simulations of an RNA structural motif suggest that, at the concentration typically used in SHAPE experiments, cooperative binding would lead to a measurable concentration-dependent reactivity.
We also provide a qualitative validation of this statement by analyzing a new set of experiments collected at different reagent concentrations.
 
\end{abstract}

\section{Introduction}

\begin{figure}
  \centering
  \includegraphics[scale=0.7]{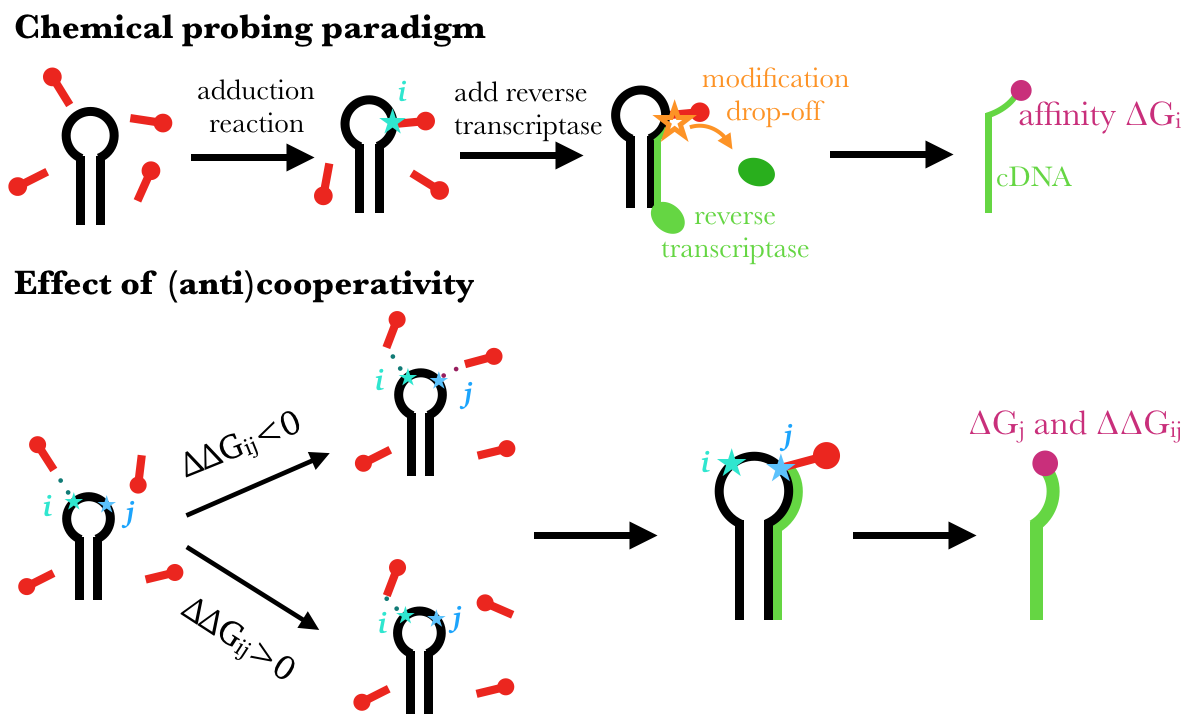}
  \caption{Chemical probing paradigm and effects of cooperativity.
In chemical probing experiments (upper panel), RNA is treated with a reagent that binds covalently.
Binding is assumed to be related to a structural determinant that depends on the specific reagent.
Reverse transcription or other techniques are then used to detect which nucleotides were reactive and thus infer structural properties of the probed motif.
Cooperativity or anti-cooperativity effects might impact observed reactivities (lower panel).
In particular, when experiments are performed at a finite reagent concentration,
a non-linear dependence of reactivity on reagent concentration is possible.
We notice that chemical binding is not required for this effect to be visible.
Even in a single hit kinetics approximation, where a single adduction per RNA molecule is observed,
multiple reagent copies might physically interact with each other and with RNA,
acting as small molecular crowders perturbing its structural dynamics.
}
  \label{fgr:cpparadigm}
\end{figure}

Chemical probing experiments allow measuring RNA structure at nucleotide resolution
by adding a chemical reagent to RNA in solution and probing at which positions
adducts are formed \cite{weeks2010advances}.
A prototypical case is the selective 2$^{\prime}$-hydroxyl acylation analyzed by primer extension (SHAPE) technique \cite{merino2005rna},
where reagents bind to the hydroxyl group of flexible nucleotides \cite{weeks2011exploring}.
This information can then be used to improve the performance of RNA structure prediction methods
(see, e.g., Refs.~\citenum{deigan2009pnas,rice2014rna,lorenz2016shape,calonaci2020nargab,saaidi2020ipanemap,cao2021characteristic,de2021progress}).
Chemical probing of small RNA molecules is usually performed
in conditions that lead to the single-hit kinetics regime, where a single adduction per RNA molecule
is formed on average  \cite{aviran2011modeling}, so that the typical spacing between adducts is on the order of a few tens of nucleotides at least.
However, it is important to note that adduction requires a prior reversible physical binding followed by an irreversible
chemical reaction. Even when the number of adductions per RNA molecule can be empirically verified, this cannot rule out
a larger number of physical binding events in the proximity of the adduction site, potentially altering RNA dynamics
and influencing the adduction rate.
These physical binding events can be considered as a form of small-molecule crowding \cite{nakano2014effects}.
Possible cooperative or anti-cooperative effects (see Fig.~\ref{fgr:cpparadigm}) might lead to unexpected
concentration-dependent reactivities.

\begin{figure}
\centering
\includegraphics[scale=0.7]{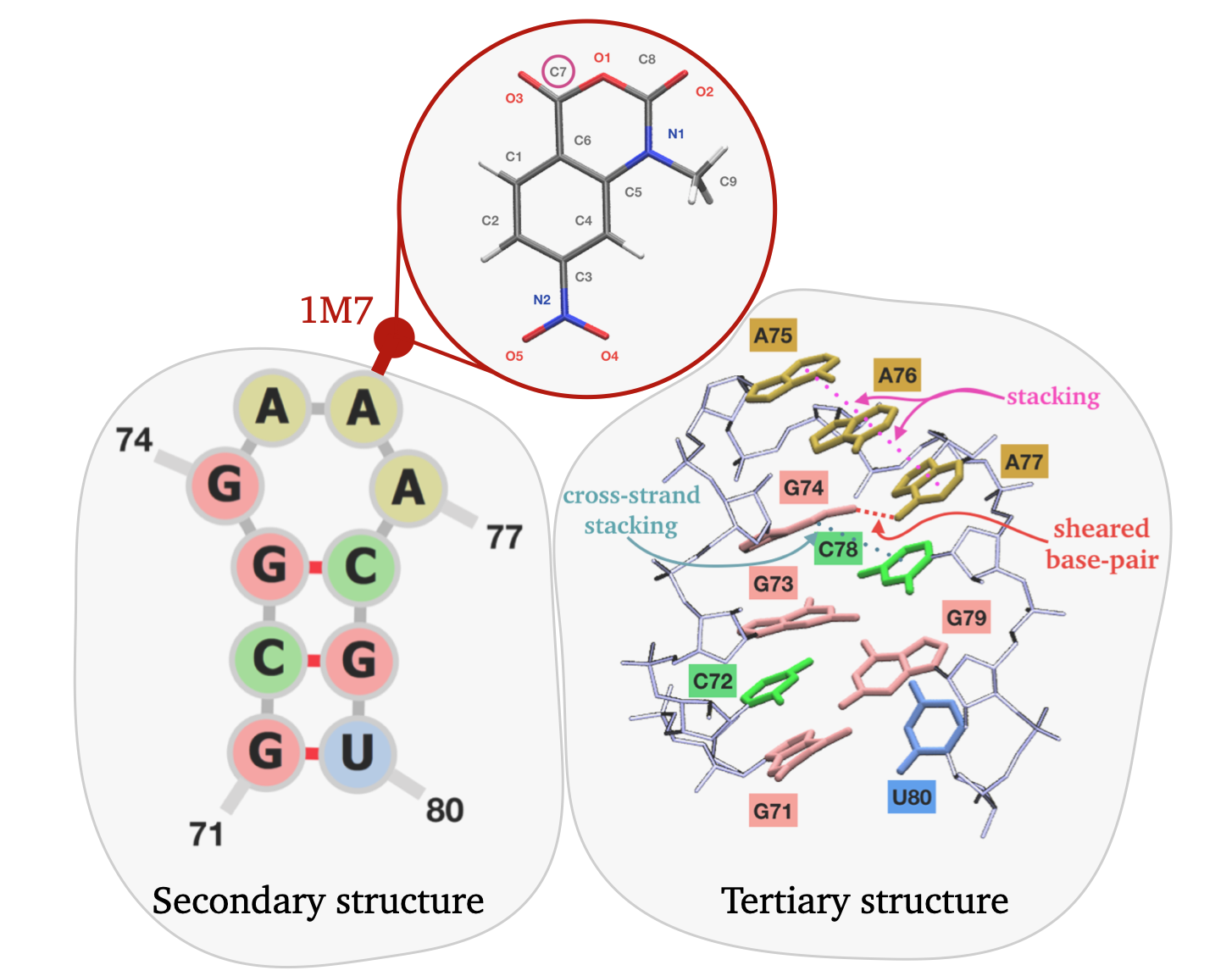}
\caption{The gcgGAAAcgu tetraloop extracted from PDB 2GIS and the chemical probing reagent 1M7.
	 Atom names of the parametrized 1M7 reagent are indicated, with the reactive site
	 C7 circled in red. In the tertiary structure 
	 representation, nucleobases are shown as thick sticks and colored consistently 
	 with the secondary structure representation; non-canonical contacts are also highlighted.}
\label{fgr:molecules}
\end{figure}

Atomistic molecular dynamics (MD) simulations give direct access to RNA dynamics \cite{sponer2018rna} and has been used to characterize
RNA flexibility and correlate it with SHAPE reactivity \cite{pinamonti2015elastic,hurst2018quantitative,mlynsky2018molecular,frezza2019interplay,hurst2021sieving}.
In some of these works, MD simulations
have been used to explicitly characterize the physical binding
of SHAPE reagents to RNA in the infinite dilution limit, where a single reagent molecule is present \cite{mlynsky2018molecular,hurst2021sieving}.
In principle, MD simulations with multiple copies of the reagent might help identifying (anti)cooperative effects
at the typical experimental concentrations.
In order to access to concentration-dependent effects, however, one should perform simulations with unrealistically large boxes
or, better, at constant chemical potential, where the number of copies of the reagent varies according to its concentration in a virtually infinite reservoir \cite{frenk-smit02book}.
Constant chemical potential simulations are usually performed using Monte Carlo techniques,\cite{papadopoulou1993molecular} which are inefficient
if a bulky reagent (see Fig.~\ref{fgr:molecules}) is to be inserted in a condensed phase.
These difficulties can be alleviated using an oscillating chemical potential \cite{lakkaraju2014sampling}, that however introduces some additional approximation,
or using nonequilibrium candidate Monte Carlo \cite{melling2022enhanced}.
These Monte Carlo methods typically require specifically modified MD codes.
Alternatively, a dedicated
region of the box can be used as a reservoir and a position-dependent potential can be added modulating  the number of copies in 
the analyzed region using adaptive-resolution \cite{wang2013grand} or constant constant-chemical-potential \cite{perego2015molecular} MD simulations.
The adaptive-resolution method is not available in general purpose MD engines, whereas constant-chemical-potential MD is
directly compatible with most simulation software via plugins such as PLUMED \cite{tribello2014plumed}.
However, both these methods require parameters such as the size and shape of the transition and reservoir regions and the form of the bias potential to be chosen in advance.

In this work, we use MD simulations to investigate (anti)cooperative effects in the physical binding of a SHAPE reagent to a typical
RNA structural motif.
We introduce an approach to grand-canonical averaging that is based on a maximum-likelihood procedure
used to analyze a set of simulations performed at a constant number of copies of the reagent molecule.
The analysis requires the solution of a self-consistent set of equations similar to those employed in weighted-histogram analysis method \cite{ferrenberg1989optimized,kumar1992weighted}
or in multistate Bennett acceptance ratio estimations \cite{shirts2008statistically}.
Importantly, the analysis is done as an \emph{a posteriori} reweighting, so that it allows to choose and optimize the reservoir region
after the simulations have been performed, computing weights to be associated to each of the simulated snapshots.
In addition, the chemical potential or, equivalently, the concentration of reagent molecules in the buffer, can be chosen \emph{a posteriori},
thus allowing for the straightforward calculation of concentration-dependent properties from a single set of simulations.
The method is then applied to compute the concentration-dependent physical binding affinity of a SHAPE reagent on an RNA tetraloop
(see Fig.~\ref{fgr:molecules}).
Interestingly, we predict that nucleotides in the loop undergo cooperative reagent binding at the typical experimental concentrations.
Experimental data supporting the existence of cooperative effects in RNA tetraloops are also reported.
Our observation opens the way to a new dimension in the interpretation of chemical probing data, where concentration-dependent
results might be used to identify specific structural motifs.

\section{Methods}
\subsection{Grand Canonical Reweighting of Molecular Dynamics}
In order to describe physical situations where the number of particles is varying, the grand-canonical ensemble is
necessary. In this ensemble the fixed quantities are the chemical potential $\mu$, that controls the fluctuations in 
the number of particles, the volume $V$ of the system and the temperature $T$. The ensemble can be represented as a 
canonical ensemble coupled to a particle reservoir that can gain or lose particles without appreciably changing $\mu$.
We consider the case where the chemical potential of a single species is controlled,
whereas all the other species are simulated at constant number of particles.
We then introduce a procedure that can be used to obtain grand-canonical averages from the combination of a set
$\mathcal{S}$ of $N_{max}$ 
independent simulations, each with a different fixed number of those particles whose chemical potential is controlled, $N \in\{1,\dots,N_{max}\}$.
The simulation box is divided into two sub-regions $A$ and $B$ that are assumed to be sufficiently decoupled.
In the $N$th simulation, which is run in the canonical ensemble, the probability to observe $k$ particles in region $A/B$ is
\begin{equation}
P^N_{A/B}\left(k\right)\propto\Omega_A\left(k\right)\Omega_B\left(N-k\right)~.
\end{equation}
Here, $\Omega_A$ and $\Omega_B$ are the canonical partition functions associated to regions $A$ and $B$, respectively.
We then define the count matrix $\mathbf{t}=\{t_{Nk}\}$ which reports, for the trajectory with $N$ copies of the particles,
how many frames were seen with exactly $k$ particles in region $A$ and $N-k$ particles in region $B$.
We notice that this is a triangular matrix, since cases where $k>N$ are impossible by construction.
The probability to observe such a matrix can be computed as the probability to generate each of the corresponding
frames and is equal to:
\begin{equation}
P(\mathbf{t}) \propto\prod\limits_{N=1}^{N_{max}}\prod\limits_{k=0}^{N}\left(c_N\Omega_A(k)\Omega_B(N-k)\right)^{t_{Nk}}.
\end{equation}
Here, the normalization coefficients $\{c_N\}$ are required to ensure that, at fixed $N$, the sum of the probabilities
$P^N_{A/B}\left(k\right)$ over $k$ is equal to one.
The maximum-likelihood (ML) estimation of $\Omega_A$ and $\Omega_B$ is obtained by minimizing the negative log-likelihood
$-\log P(\mathbf{t})$. By using the Lagrange multiplier methods to include the normalization constraint mentioned above,
one obtains the following Lagrangian function
\begin{equation}
\label{eq:likelihood}
	\mathcal{L}=-\sum\limits_{N=1}^{N_{max}}\sum\limits_{k=0}^{N} t_{Nk}\log\left(c_N\Omega_A(k)\Omega_B(N-k)\right)-\sum_N \lambda_N\left(\sum\limits_k c_N\Omega_A(k)\Omega_B(N-k)-1\right)~,
\end{equation}
where $\{\lambda_N\}$ are $N_{max}$ Lagrangian multipliers.
        The notation can be simplified by defining: $A_k=\sum_Nt_{Nk}$, counting the number of times that,
        in the whole set of $N_{max}$ trajectories, a particle was found in region $A$;
        $B_k=\sum_Nt_{N,N-k}$, counting the equivalent number for region $B$;
        and $L_N=\sum_k t_{Nk}$, the total number of frames accumulated in the trajectory with $N$ particles.
In Section~\ref{sec:likelihood-maximization} we show that minimizing Eq.~\ref{eq:likelihood} as a function $\Omega_A$ and $\Omega_B$ leads to the following coupled equations:
\begin{equation}
\begin{aligned}
&\Omega_A(k)=\frac{A_k}{\sum_N L_N c_N \Omega_B(N-k)}\\
&\Omega_B(k)=\frac{B_k}{\sum_N L_N c_N \Omega_A(N-k)}~.
\end{aligned}
\end{equation}
These equations can be solved iteratively through the procedure reported in Algorithm~\ref{gcsolve}.
Since we made no assumption on the length of each trajectory $\{L_N\}$, the method can be straightforwardly used also when the minimum number
of simulated copies of the controlled particles is greater than 1 or when some simulations are missing, by just setting some of the elements of $L_N$ to zero.
Once the ML estimates of $\Omega_{A/B}$ have been obtained, they can be directly plugged
in the grand-canonical probability of observing molecules in regions $A/B$, which is defined as
\begin{equation}
P^{GC}_{A/B}\left(N_{A/B}\right)\propto\Omega_{A/B}\left(N_{A/B}\right)e^{-\mu N_{A/B}/RT}
\end{equation}
This expression can be then used to compute the grand-canonical average of the number of particles in both regions $A$ and $B$,
at a fixed value of chemical potential $\mu$.
\begin{equation}\label{nanb}
\langle N_{A/B}\rangle_{GC}=\sum\limits^{N_{max}}_{k=0}k\cdot P^{GC}_{A/B}\left(k\right)
                           =\frac{\sum\limits^{N_{max}}_{k=0}k\cdot\Omega_{A/B}(k)e^{-\mu k/RT}}{\sum\limits_{k}^{N_{max}}\Omega_{A/B}(k)e^{-\mu k/RT}}
\end{equation}
Equation \ref{nanb} provides a connection between the concentration in the experimental buffer and
the chemical potential  $\mu$.
Specifically, one can use the bisection method reported in Algorithm~\ref{musolve} to obtain $\mu$ corresponding to the desired concentration in region $B$ (see Section~\ref{sec:fixing-mu-of-n}).

Once $\mu$ has been obtained, grand-canonical averages in region $A$ can be obtained by weighting frame $i$
with a factor
\begin{equation}
w_i\propto \frac{\Omega_A(k_i)e^{-\mu k_i/RT}}{A_{k_i}}
\end{equation}
where $k_i$ is the number of copies of the molecule in region $A$ in that frame.
These weights, and the relationship between the concentration in region $B$ and the chemical potential $\mu$ (Eq.~\ref{nanb}),
can be used to obtain arbitrary grand-canonical ensemble averages as smooth continuous functions of the concentration.

In short, our method is composed of the following steps:
(a) a number of simulations are performed with different number of copies of the particles and concatenated;
(b) histograms counting how many times particles are present in region $A$ and $B$ are computed;
(c) these histograms are used to compute the canonical partition functions $\Omega_A$ and $\Omega_B$;
(d) $\Omega_B$ is used to calculate which is the chemical potential $\mu$ corresponding to a given concentration;
(e) $\mu$ and $\Omega_A$ are used to compute the weight associated to each of the frames of the initial concatenated
trajectory.

\subsection{Lattice model}
We test the method on a lattice space divided in two regions, $A$ and $B$, containing $S_A$ and $S_B$ sites, respectively. 
Sites are then populated with a number up to $N_{max}$ of particles that interact only through mutual exclusion: 
a site cannot be occupied by more than one particle. Two scenarios are tested: a purely entropic lattice, 
in which the free energy depends only on the number of possible configurations of the particles occupying the
$S=S_A+S_B$ sites; and a lattice with one stabilizing site in region $A$ that brings in a negative contribution to the 
free energy.
The latter is supposed to mimic the situation where reagent molecules can bind to an RNA molecule that is
located in region $A$. For more details, see Section~\ref{sec:ci:lattice}.

\subsection{GAAA tetraloop of SAM-I riboswitch} We then apply the introduced method to an RNA GNRA tetraloop (here N is any nucleotide and R is G or A), as this type of structural
motif has some well-established properties: it presents (a) highly stable secondary structure \cite{heus1991structural} along with (b) rich dynamics involving 
multiple non-canonical contacts \cite{sorin2010nar,hall2015mighty}, that could lead to significant structural changes when in contact with SHAPE reagents; 
noticeably, (c) in SHAPE experiments the GNRA tetraloop presents a typical reactivity pattern \cite{mlynsky2018molecular}.
We simulate a single loop motif rather than duplexes or larger structures in order to keep computational costs low,
under the hypothesis that long-range effects are negligible. We expect this hypothesis to be reasonable as there is no
evidence of conformational rearrangements due to interaction with SHAPE reagents, rather than at a local scale \cite{mlynsky2018molecular}. 
The gcgGAAAcgu tetraloop is taken from the annotated structure of SAM-I riboswitch, that can be found in the PDB entry 2GIS \cite{montange2006structure}.
A representation of the resulting construct is shown in Fig. \ref{fgr:molecules}.
The stretch obtained in this way consists in a sequence of three base pairs, namely G71-U80, C72-G79 and G73-C78, plus
the tetraloop under study: G74-A-A-A77.
The initial conformation for this molecule is obtained by extracting the coordinates of the corresponding
atoms from the PDB 2GIS entry.
The closing base-pair of the sequence (G71-U80) is observed to unpair in preliminary simulations where a larger number of reagents is used. 
Since the calculation is meant to be representative of a GNRA tetraloop embedded in a longer RNA molecule,
a harmonic restraint is applied to the 
hydrogen bonds between 71G/O6 and 80U/N3 and between 71G/N1 and 80U/O2, and these bases are excluded from analysis 
of reactivity and cooperativity to minimize terminal effects.
RNA is parametrized according to the AMBER force field \cite{cornell1995second,perez2007refinement,zgarbova2011refinement}.

\subsection{Parametrization of 1-methyl-7-nitroisatoic anhydride (1M7)} 1M7 is an efficient reagent used for SHAPE probing \cite{mortimer2007jacs}. The molecule is parametrized according to the general Amber force field (GAFF) 
\cite{wang2004jcc,wang2006jmgm} for organic molecules using the Antechamber and parmchk tools implemented in Ambertools \cite{case2005jcc}.
The 1M7 probe structure is generated through the Maestro interface of the
Schr{\"o}dinger suite \cite{maestro}. The Gaussian 16 package is then employed for geometrical optimization and calculation
of the electrostatic potential of the probe, using the B3LYP hybrid functional method with $6-31G^*$ basis set.
Partial charges are then calculated using the RESP method \cite{cornell1993jacs} as implemented in Antechamber.
The resulting charges, that sum up to 0 as 1M7 is overall neutral, are reported in Table~\ref{par1m7}.
The resulting Amber potential is then converted to the GROMACS implementation \cite{abraham2015soft}, using acpype
\cite{bernardi2019soft}. 
The optimized structure of 1M7 is reported in Fig. \ref{fgr:molecules}.
 
\subsection{Simulation protocol}
In order to sample a range of different concentrations of 1M7, $N_{max}=19$ independent simulations are set up, each
featuring a fixed number of probes, from $N=1$ to $N=N_{max}$. For each of them, the center of mass of the tetraloop is 
taken as origin of the reference frame. A rhombic dodecahedron simulation box is placed at a distance of $\SI{3}{\nano\meter}$ 
from the tetraloop. It is important to place the box at this step, before inserting the 1M7 probes, in order to preserve the 
volume across the simulations with different $N$. Reagents are placed at random points at equal distance from the tetraloop 
and with random orientation. In particular, the first probe is placed at a random point on the surface of a sphere, centered on
the tetraloop and with radius equal to the radius of gyration of the tetraloop plus $\SI{2}{\nano\meter}$. The probe is then 
rotated about its center of mass by a random angle. A check on the distances between every atom pair is made in order to 
avoid clashes: if one of the atoms of the inserted probe is at a distance lower than $\SI{5}{\angstrom}$ from any other atom, 
the insertion is rejected and another point and orientation are generated. For each of the remaining $N-1$ probes the insertion 
procedure is repeated. Examples of the resulting conformation are represented in Fig.~\ref{start_conf} for $N=5$ and $N=16$.
The resulting complexes are solvated using the OPC water model \cite{izadi2014jpcl} and sodium counterions are added
to neutralize the system \cite{joung2008determination}.
For each complex, the potential energy is minimized in order to relax the structures and remove
possible clashes and incorrect geometries, through 50000 steps of steepest descent algorithm. The minimization is
followed by NVT equilibration of $\SI{1}{\nano\second}$ up to a temperature $T=\SI{300}{\kelvin}$, and NPT equilibration at the
same temperature, pressure $P=\SI{1}{\bar}$ for another $\SI{1}{\nano\second}$ using a Parrinello-Rahman barostat
\cite{parrinello1981jap}. A cutoff of $\SI{10}{\angstrom}$ and the particle-mesh Ewald (PME) method \cite{essmann1995jcp} are used for
computing short-range interactions and long-range interactions, respectively. Temperature is controlled using the stochastic velocity rescaling
thermostat \cite{bussi2007jcp}. Equilibration is run with a time step of $\SI{2}{\femto\second}$ with bonds involving hydrogens constrained via the
LINCS algorithm \cite{hess1997jcc}. Production runs are then carried out in the NPT ensemble at $T=\SI{300}{\kelvin}$ and $P=\SI{1}{\bar}$.
Plain MD simulations are performed using version 2018.5 of the GROMACS software \cite{abraham2015soft}.

\subsection{Statistical uncertainties}

To compute statistical uncertainties, we rely on a Bayesian bootstrap procedure \cite{rubin1981bayesian}
where each entire trajectory is treated as a single data point.
Specifically, at each bootstrap iteration (total $N=$10000 iterations) we extract 19 weights from a Dirichlet distribution
and use the resulting weighted trajectories to (a) estimate the canonical partition functions $\Omega_A$ and $\Omega_B$,
(b) compute the chemical potential $\mu$ corresponding to the desired concentration in region $B$ and (c)
use the resulting weights to compute the observable of interest.
Given that trajectories are independent of each other,
and at variance with standard block analysis \cite{flyvbjerg1989error},
this estimate of the uncertainty is not subject to errors due to correlation between data points.

\subsection{Experimental methods}

DNA template corresponding to the GNRA tetraloop containing RNAs used in this study
(PDB entries
2GIS\cite{montange2006structure},
 1KXK\cite{zhang2002structural},
 1SCL\cite{szewczak1993conformation},
 1CQ5\cite{schmitz1999structure},
 and 2GV4\cite{zoll2007breaking})
with 5$^{\prime}$ and 3$^{\prime}$ SHAPE cassettes \cite{wilkinson2006selective} and the T7 promoter sequence was ordered from Eurofins Genomics. The RNA was transcribed and purified as previously described \cite{calonaci2020nargab}. SHAPE experiments were carried out with the 1M7 adduct at three final concentrations (12.5 mM, 6.5 mM, and 3.25 mM) and subsequent analysis of the concentration series was carried out as previously described \cite{calonaci2020nargab}.

\section{Results}

\subsection{Lattice model}

In order to highlight the potential limitations of the simulation and reweighting protocol
we used our method to reconstruct the grand-canonical distributions for a lattice model.
Results are presented in the supporting information, Section~\ref{sec:ci:lattice}, and highlight the main limitation of the method,
namely the fact that only concentrations that correspond to the number of particles in the set of
analyzed simulations can be correctly reproduced. In addition, the model can be used to study the impact of statistical
sampling errors on the estimated distributions.

\subsection{Molecular dynamics simulations of SHAPE reagents}

In order to estimate the reactivity profile and cooperativity matrix of the gcgGAAAcgu tetraloop at different concentrations 
of 1M7, we first divide the simulation space into two regions: the binding region $A$ is 
spherical, centered at the center of mass of the RNA motif with a fixed radius $r_A$; the rest of the simulation space is defined 
as the buffer region $B$. In the binding region, the reagent copies  are in proximity of the tetraloop and can form a relatively 
stable bound state, preliminary to the formation of the covalent bond that is not modeled here. In the buffer region, there are no direct interactions between reagent copies
and RNA, and the formation of a bound state is not possible, as $r_A$ is beyond the range of distances
for binding. $N_{max}=19$ trajectories are collected, each featuring $N$ reagent copies with $N\in\left[1,\dots,N_{max}\right]$.
Every trajectory contains $10^{5}$ frames, corresponding to a total simulation length of $\SI{1}{\micro\second}$ per trajectory. %
From the entire set of trajectories, we compute the number of times that each pair of nucleotides $i$ and $j$ in the tetraloop 
are in one of four possible pairwise binding states: both unbound, both bound to 
two different reagent copies, or only one of the two nucleotides bound to a reagent copy.
We define binding between a nucleotide and a reagent copy to occur whenever the following two conditions
are satisfied: a) the nucleotide is the nearest one to the probe, and b) the distance between the nucleotide and the probe is
less than a certain threshold. For both conditions, we measure the distance by considering the atoms involved in the chemical reaction,
namely the O$2^\prime$ atom of the nucleotide and the C$7$ atom of the reactive carbonyl of 1M7.
We set this threshold to $r_{th}=\SI{3.5}{\angstrom}$, consistently with Ref.~\citenum{mlynsky2018molecular}.

By accumulating statistics on the $N_{max}$ trajectories and using the introduced grand-canonical reweighting,
we can estimate the partition functions $\Omega_A$ and $\Omega_B$, the value of the chemical potential $\mu$ that correspond to the target reagent
concentration, and the probability for one or two nucleotides to be bound to a reagent in the grand-canonical ensemble.
Typically, SHAPE experiments using 1M7 as a probe are carried out at reagent concentrations ranging from $0.1$ to 
$\SI{100}{\milli\Molar}$ \cite{mortimer2007jacs}.
Using a dodecahedral simulation box of volume $V_{box}\approx\SI{400}{\nano\meter}^3$, the radius of the binding region
fixed at $r_A=\SI{3}{\nano\meter}$ and $N_{max}=19$ maximum number of reagent copies in the collected trajectories
we can only reproduce reagent concentrations that are below a threshold of 
$C_{max}=\frac{N_{max}}{N_A\left(V_{box}-4/3\pi r_A^3\right)}\approx\SI{10}{\milli\Molar}$.
Given $\mu$ and 
$\Omega_A$, one can obtain the weights for computing averages in the grand-canonical ensemble,
which are denoted as $w(N_A)$ since, for each frame, the weight only depends on the number of copies of the reagent seen in region $A$.

\subsection{Concentration-dependent reactivities}

\begin{figure}
  \centering
  \includegraphics[scale=0.7]{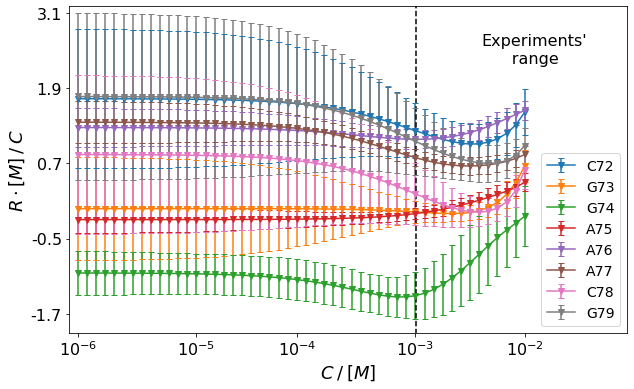}
  \caption{Reactivity, computed as the probability of each nucleotide to be physically bound to a SHAPE reagent, shown as a function of reagent concentration.
Concentrations are reported in molar units,
and the vertical dashed line denotes a typical lower bound for experimental concentrations (1mM).
}
  \label{fgr:rvsc}
\end{figure}
For each nucleotide, we estimate the reactivity $R_i$ from the frequency with which it is observed in a bound state with any one of 
the reagent copies, and we average it in the grand-canonical ensemble, using the weights $w(N_A)$.
In theory, the relation between the reactivity of a nucleotide and the concentration of reagent can be decomposed in the sum 
of terms representing the effect of a) the number of available reagent copies, b) the effect of a bound nucleotide 
on the binding probability of another nucleotide, due to (positive or negative) cooperativity, and c) higher order relations involving more than 
two nucleotides. The first term is proportional to the concentration while higher-order terms depend on higher powers of the concentration.  
The dependency on reagent concentration of the binding affinities that we obtain for the simulated RNA motif is 
represented in Fig.~\ref{fgr:rvsc}. At sufficiently low concentrations, the ratio $R/C$ between reactivity and concentration saturates 
to a constant, consistently with the expected linear relationship. As reagent concentration is increased, and specifically 
for $C>\SI{e-3}{\Molar}$,
higher-order contributions start to emerge significantly, as some reactivities show a 
non-linear dependency on concentration. 
In principle, one nucleotide can exhibit positive cooperativity with some nucleotides and negative cooperativity with others.
We attribute a super-linear relation between reactivity and reagent concentration to predominantly cooperative behavior, 
as binding affinity increases more than proportionally to the number of reagent copies available. As well, we interpret 
a sub-linear dependency as a signal of predominantly anti-cooperative behavior, and an approximately linear dependency either as absence
of cooperative effects or as positive and negative cooperativity behaviors compensating each other.
Noticeably, the range of concentrations ($10^{-3}$ to $\SI{e-2}{\Molar}$) that we identify as affected by cooperative 
effects, overlaps significantly with the range of concentrations typically adopted in experiments.
In this range, we quantify the non-linearity of reactivity as a function of concentration for each nucleotide, by fitting power laws $R_i=\alpha\cdot C^{\beta}$.
Fit parameters are reported in Table \ref{tbl::powerfit}.
Uncertainties are computed here by performing the fitting at every bootstrap iteration and computing the standard deviation of the resulting coefficients.
Although the statistical uncertainty on the individual points reported in Fig.~\ref{fgr:rvsc} is relatively high,
errors associated to different values of concentration are correlated, resulting in relatively low uncertainty in
the estimated power coefficients (Table \ref{tbl::powerfit}).
In particular, for G74 we detect the strongest super-linear dependency.
\begin{table}
\centering
\begin{tabular}{l c c}
\multicolumn{3}{l}{$R = \alpha\cdot C^\beta$}\\
\hline
Nucleotide & $\beta$ & $\alpha$\\
\hline
C72 & $1.1\pm 0.2$ & $11\pm 14$ \\
G73 & $1.3\pm 0.1$ & $ 7\pm  6$ \\
G74 & $1.6\pm 0.2$ & $19\pm 24$ \\
A75 & $1.2\pm 0.1$ & $ 4\pm  2$ \\
A76 & $1.2\pm 0.1$ & $13\pm  6$ \\
A77 & $1.0\pm 0.1$ & $ 3\pm  1$ \\
C78 & $1.1\pm 0.2$ & $ 4\pm  5$ \\
G79 & $0.9\pm 0.2$ & $ 3\pm  4$ \\
\hline
\end{tabular}
\caption{Parameters of a power law of reactivity $R$ as a function of concentration $C$, obtained by
        a least-squares linear fit of their logarithms, for each of the analyzed nucleotides of the gcgGAAAcgu tetraloop. Powers $\beta>1$ indicate
        cooperative behavior, while $\beta<1$ indicate anti-cooperativity.
        Standard errors computed using bootstrap are reported.}
\label{tbl::powerfit}
\end{table}

\subsection{Free-energy couplings}

\begin{figure}
  \centering
  \includegraphics[scale=0.4]{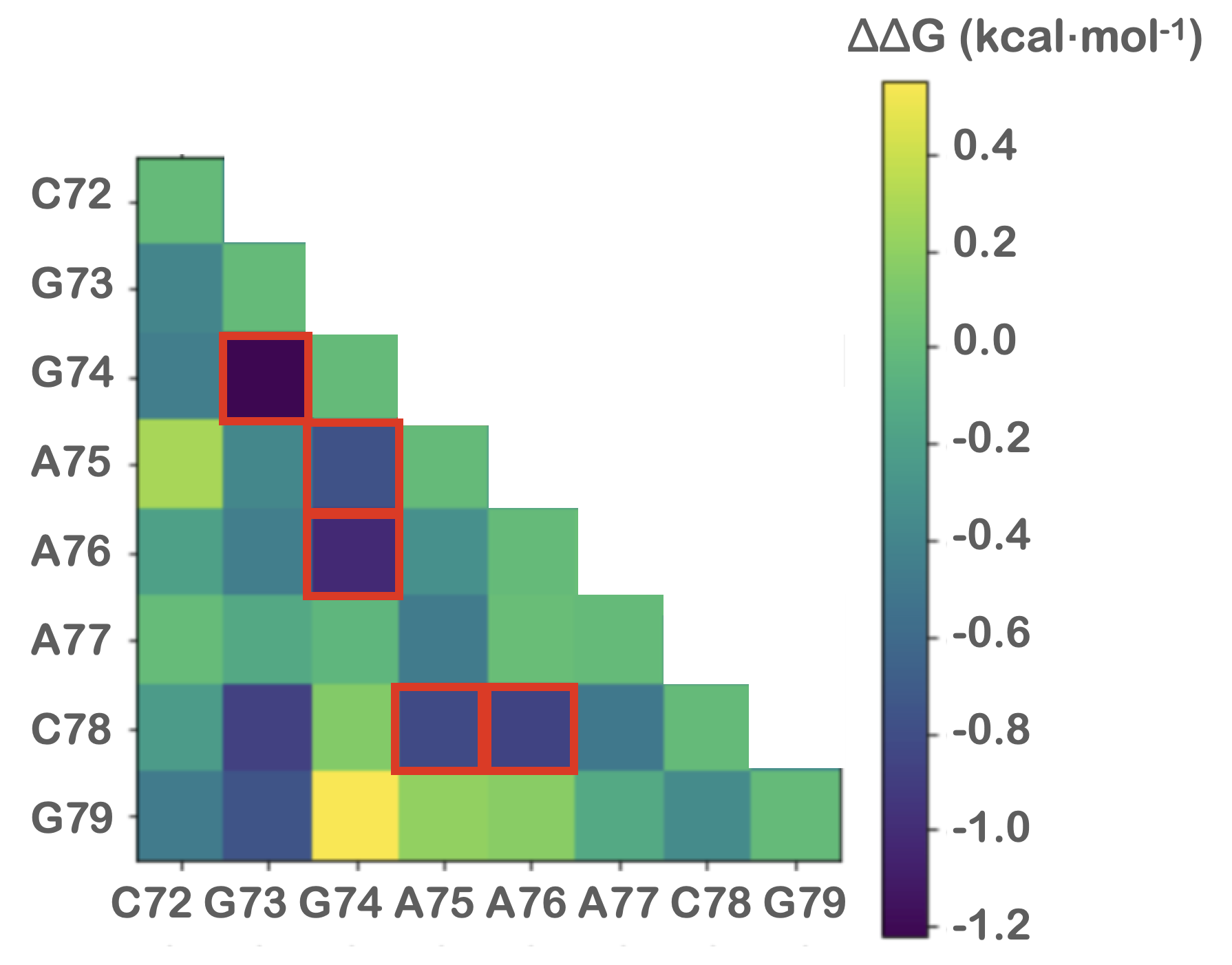}
  \caption{Cooperativity matrix $\Delta\Delta G$ at typical reagent concentration.
Pairs of nucleotides for which the cooperativity is different from zero with significance level greater than 0.01
are highlighted in red.
Anti-cooperative pairs with significance level greater than 0.01 are not observed.
}
  \label{fgr:ddg}
\end{figure}

In order to quantify the cooperativity of nucleotides in reagent binding, we rely on the free-energy coupling model \cite{forsen1995tbs}.
Negative free-energy coupling $\Delta\Delta G_{ij}<0$ means that the binding affinity of nucleotide $i$ is increased if nucleotide $j$ is
bound to a reagent copy, so they are cooperative. Vice-versa, positive $\Delta\Delta G_{ij}>0$ means they are anti-cooperative. 
From the observed events, we can compute the frequency with which two nucleotides are in the same binding states, and the frequency
with which only one of the two is bound. From the ratio between these two frequencies, we compute the free-energy coupling for each pair of 
nucleotides, reweighted in the grand-canonical ensemble. 
The estimated values of $\Delta\Delta G$ for an intermediate reagent concentration ($C=\SI{5.7}{\milli\Molar}$) among
the tested ones are reported in Fig.~\ref{fgr:ddg}.

To identify pairs of nucleotides for which the cooperativity or anti-cooperativity is significantly different from zero,
we check which fraction of the bootstrap samples return a cooperativity or anti-cooperativity larger than zero.
We set a significance level of $\alpha=0.01$. Since we deal with 28 
hypotheses simultaneously, we rely on the Benjamini-Hochberg procedure \cite{benjamini1995jrs} to keep the false discovery rate of our estimates 
at level $\alpha$. Pairs of nucleotides with significantly cooperative behavior are G73-G74, G74-A76, A76-C78
, G74-A75 and A75-C78. At the same concentration, no significant anti-cooperative behavior is identified.

\begin{figure*}
\centering
\begin{subfigure}[t]{0.33\textwidth}
        \includegraphics[scale=0.20]{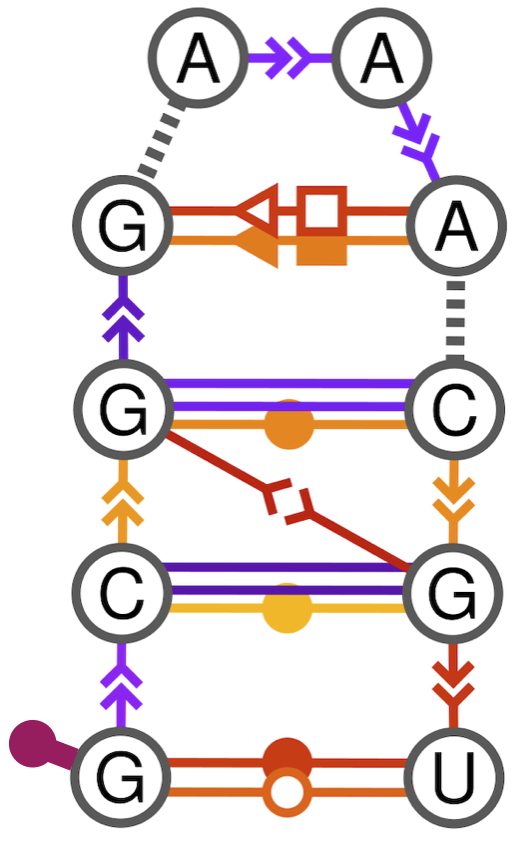}
    \caption{}
    \end{subfigure}\hfill
    \begin{subfigure}[t]{0.33\textwidth}
        \includegraphics[scale=0.20]{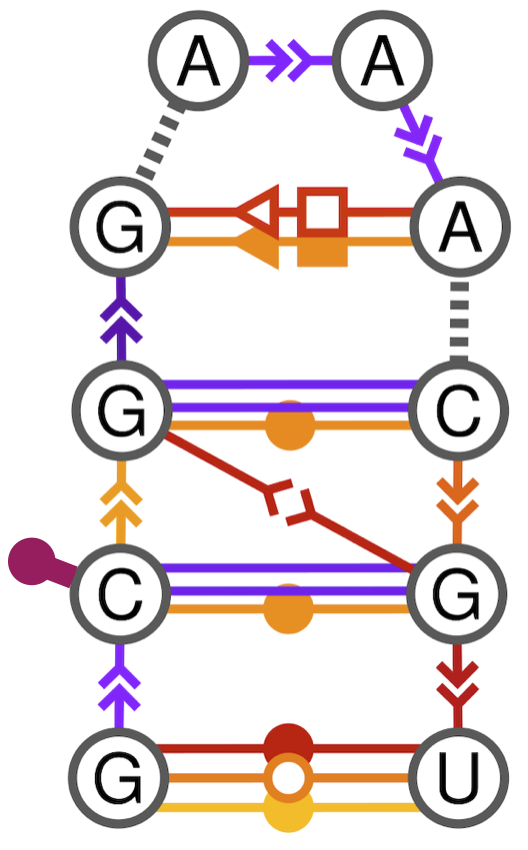}
\caption{}
    \end{subfigure}\hfill
    \begin{subfigure}[t]{0.33\textwidth}
        \includegraphics[scale=0.20]{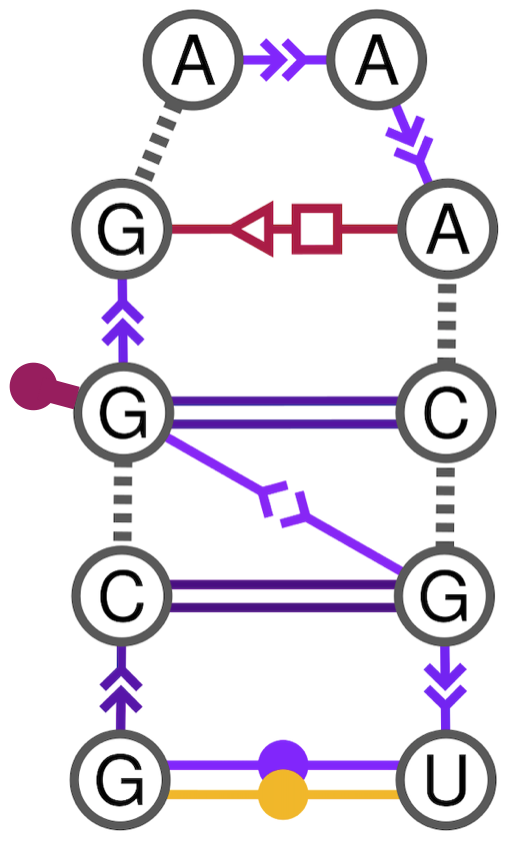}
    \caption{}
    \end{subfigure}\vfill
    \begin{subfigure}[t]{0.33\textwidth}
        \includegraphics[scale=0.20]{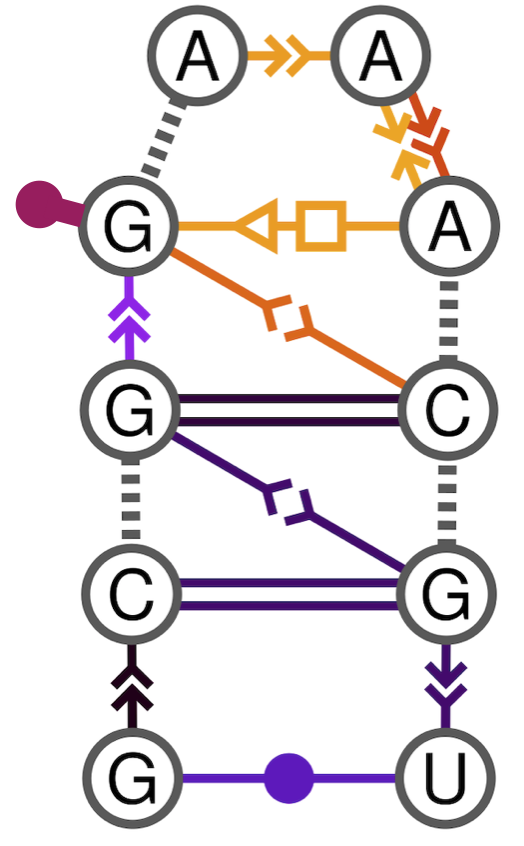}
\caption{}
    \end{subfigure}\hfill
    \begin{subfigure}[t]{0.33\textwidth}
        \includegraphics[scale=0.20]{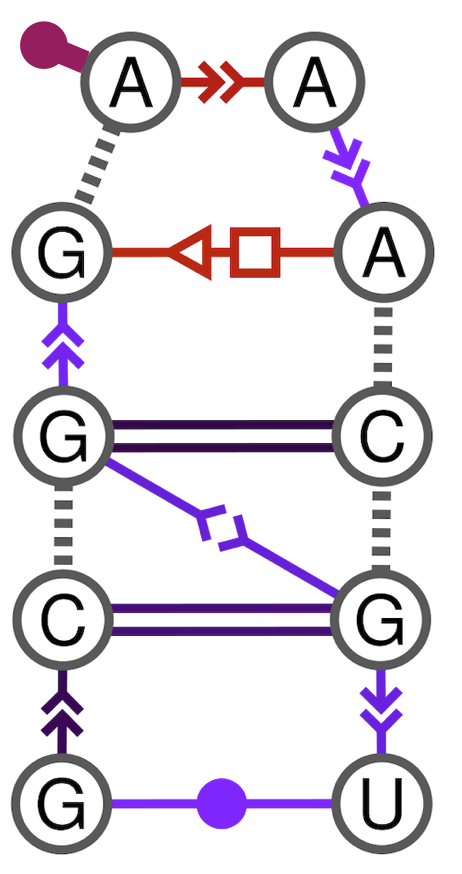}
\caption{}
    \end{subfigure}\hfill
    \begin{subfigure}[t]{0.33\textwidth}
	    \includegraphics[scale=0.20]{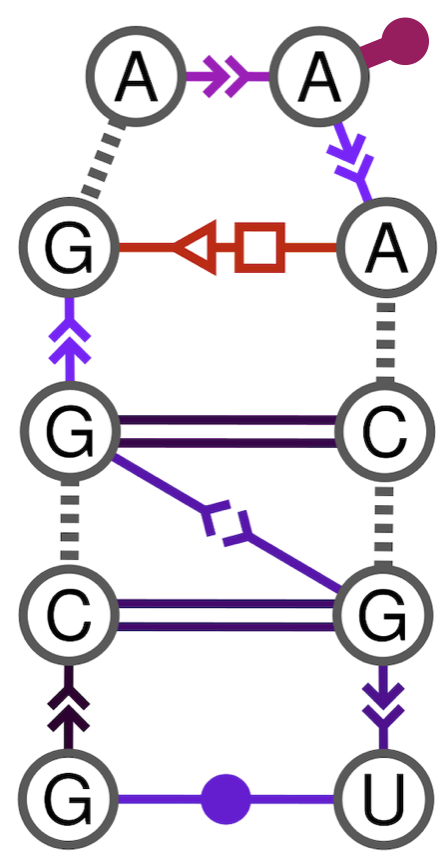}
\caption{}
    \end{subfigure}\vfill
\begin{subfigure}[t]{0.33\textwidth}
        \includegraphics[scale=0.20]{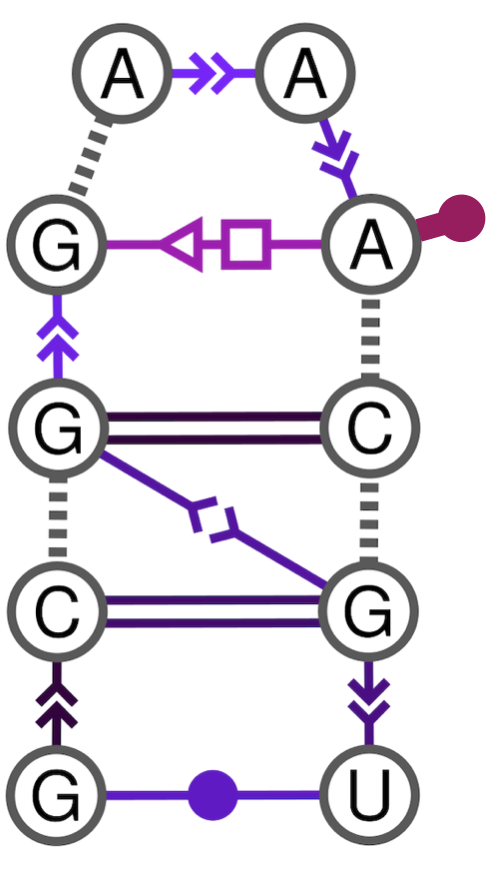}
    \caption{}
    \end{subfigure}\hfill
\begin{subfigure}[t]{0.33\textwidth}
        \includegraphics[scale=0.20]{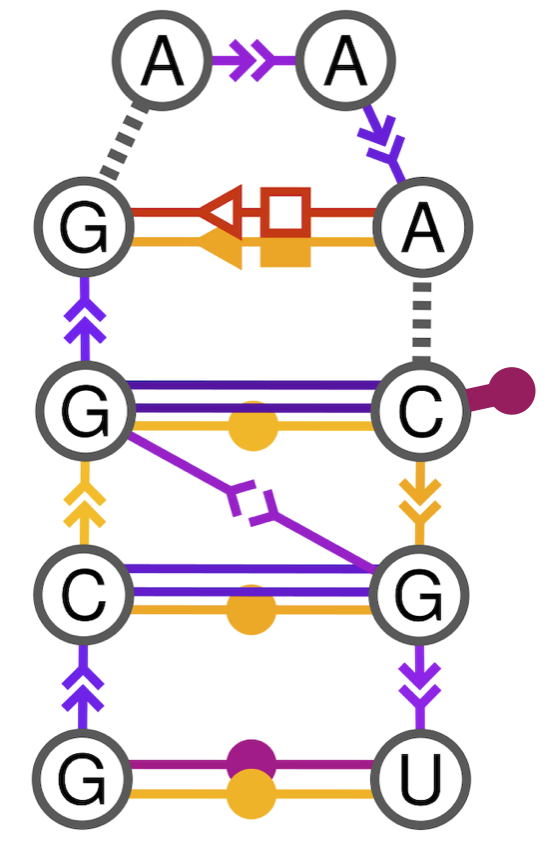}
    \caption{}
    \end{subfigure}\hfill
\begin{subfigure}[t]{0.33\textwidth}
        \includegraphics[scale=0.30]{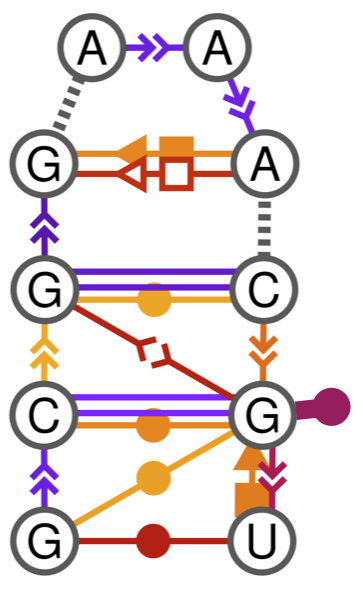}
    \caption{}
    \end{subfigure}\vfill
\begin{subfigure}[t]{0.33\textwidth}
        \includegraphics[scale=0.30]{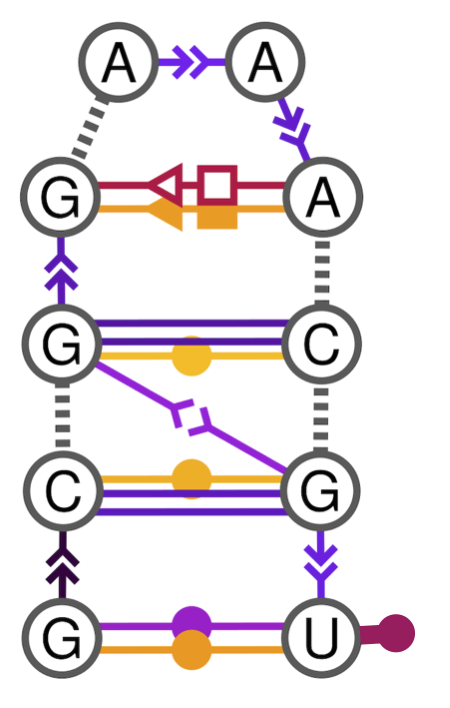}
    \caption{}
    \end{subfigure}\hfill
\begin{subfigure}[t]{0.33\textwidth}
        \includegraphics[scale=0.20]{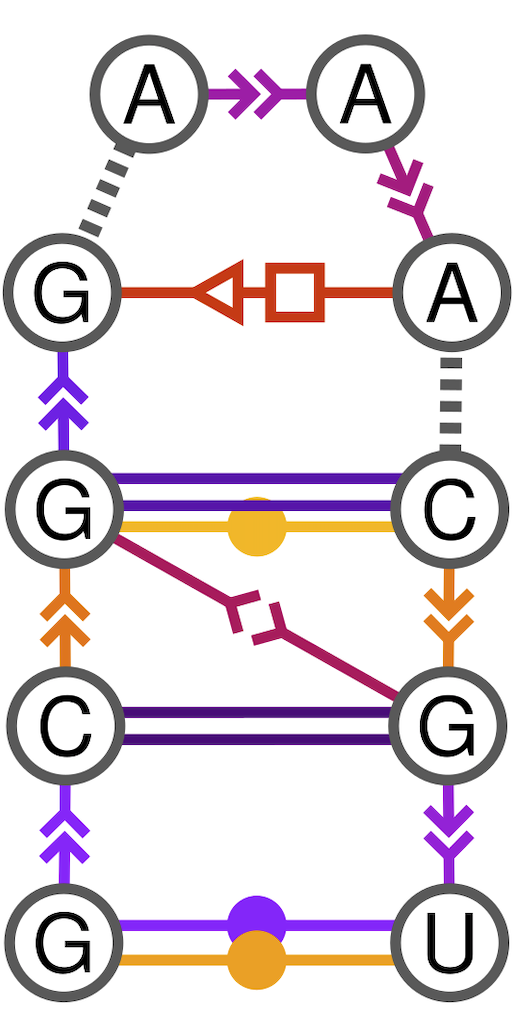}
    \caption{}
    \end{subfigure}\hfill
    \begin{subfigure}[t]{0.33\textwidth}
	\includegraphics[scale=0.22]{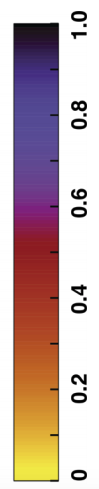}
    \end{subfigure}
	\caption{Dynamic secondary structures showing the probability of annotated interactions within grand-canonical reweighted ensembles
	constrained by individual binding of each nucleotide.  Binding is represented through a sketch of the reagent
        (in red). The bound nucleotides are (a) G71 (b) C72, (c) G73, (d) G74, (e) A75, (f) A76, (g) A77, (h) C78, (i) G79, (j) U80.
	(k) The same analysis with no constraint on binding. Base pairings are displayed using the Leontis-Westhof notation \cite{leontis2001geometric,bottaro2019barnaba}. The colormap indicates the population of each of the annotated interactions.}
\label{dynstru_single}
\end{figure*}

\subsection{Structural analysis}
\begin{figure*}
\centering
    \begin{subfigure}[t]{0.33\textwidth}
        \includegraphics[scale=0.35]{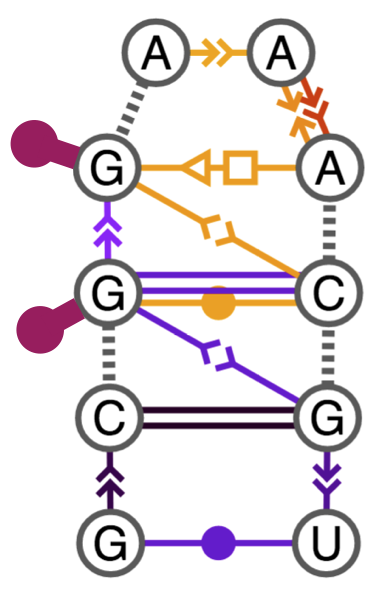}
    \caption{}
    \end{subfigure}\hfill
    \begin{subfigure}[t]{0.33\textwidth}
        \includegraphics[scale=0.35]{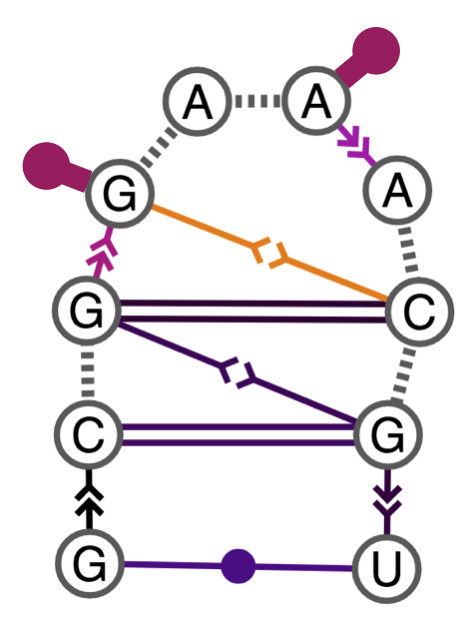}
\caption{}
    \end{subfigure}\hfill
    \begin{subfigure}[t]{0.33\textwidth}
        \includegraphics[scale=0.35]{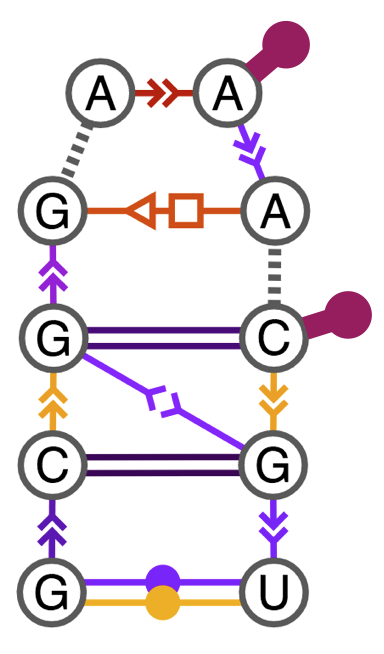}
\caption{}
    \end{subfigure}\vfill
    \begin{subfigure}[t]{0.33\textwidth}
        \includegraphics[scale=0.35]{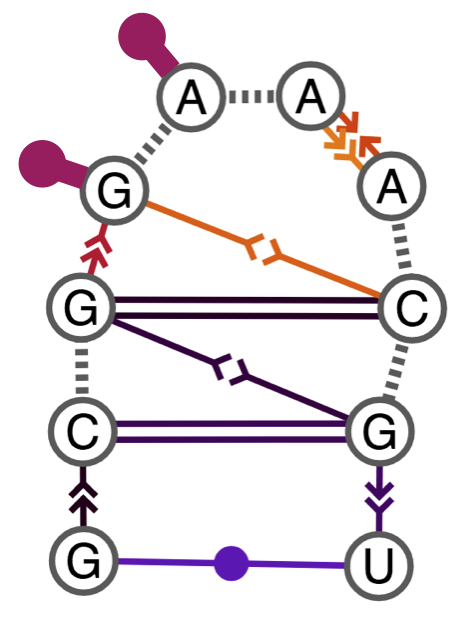}
\caption{}
    \end{subfigure}\hfill
    \begin{subfigure}[t]{0.33\textwidth}
	\includegraphics[scale=0.35]{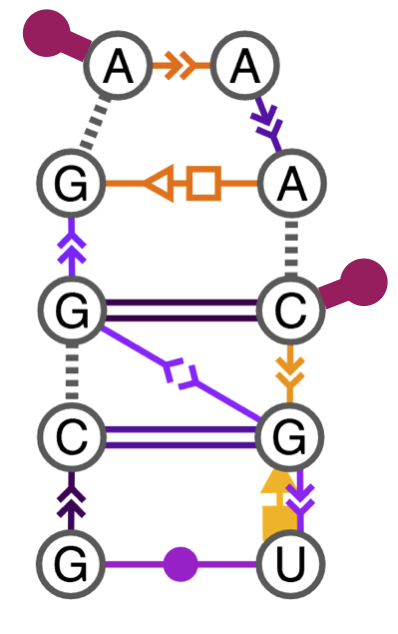}
\caption{}
    \end{subfigure}\hfill
    \begin{subfigure}[t]{0.33\textwidth}
	\includegraphics[scale=0.25]{dyn_colormap.png}
    \end{subfigure}
	\caption{Dynamic secondary structures showing the probability of annotated interactions within grand-canonical reweighted ensembles,
	constrained by simultaneous binding of cooperative pairs of nucleotides. Binding is represented through a sketch of the reagent
	(in red).
	The represented pairs are (a) G73 and G74, (b) G74 and A76, (c) A76 and C78, (d) G74 and A75, (e) A75 and C78.
        Base pairings are displayed using the Leontis-Westhof notation \cite{leontis2001geometric,bottaro2019barnaba}.
        The colormap indicates the population of each of the annotated interactions.}
\label{dynstru}
\end{figure*}

\begin{figure*}
    \begin{subfigure}[ct]{0.33\textwidth}
        \includegraphics[scale = 0.2]{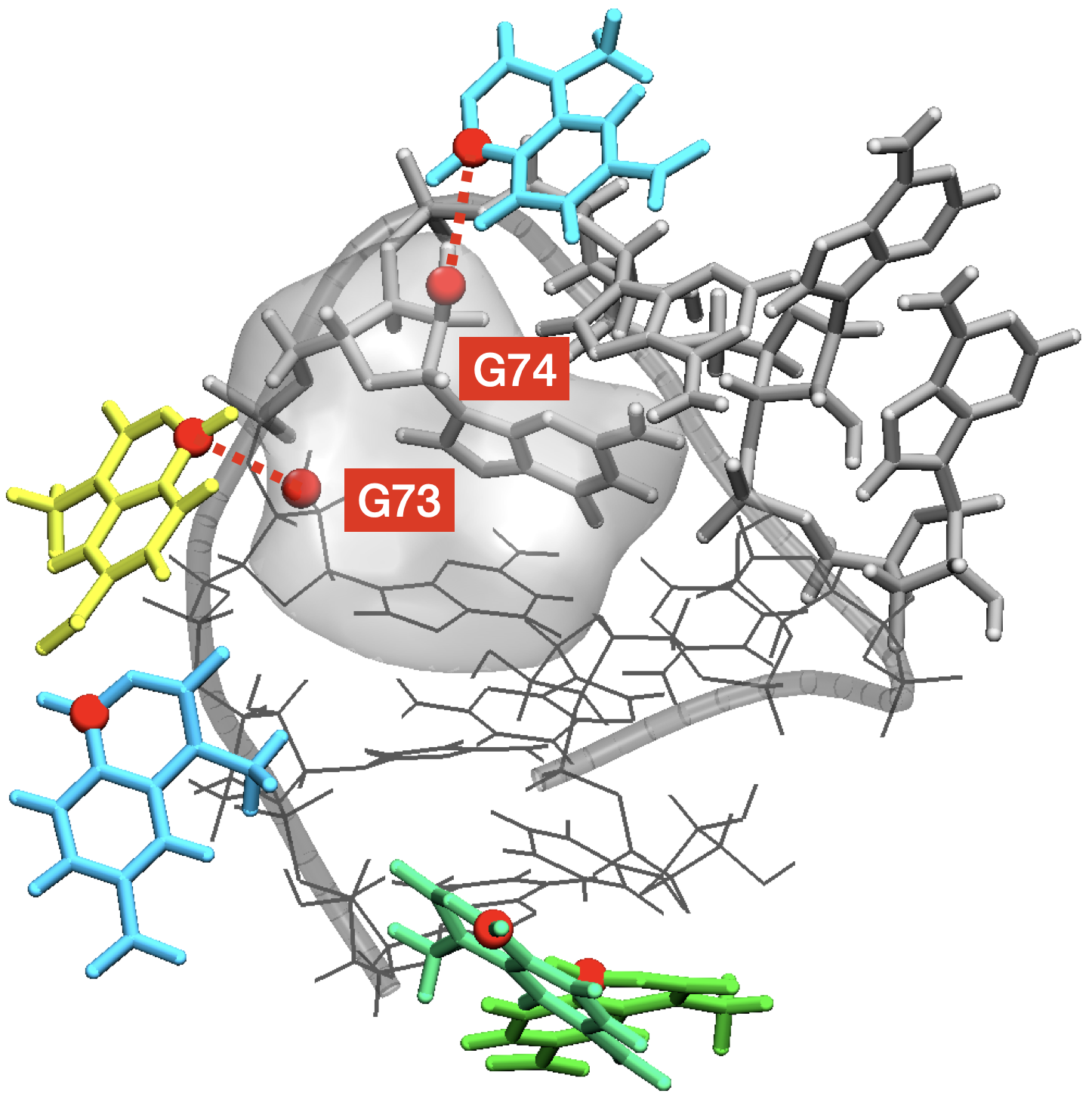}
    \caption{}
    \end{subfigure}
    \begin{subfigure}[ct]{0.33\textwidth}
        \includegraphics[scale = 0.2]{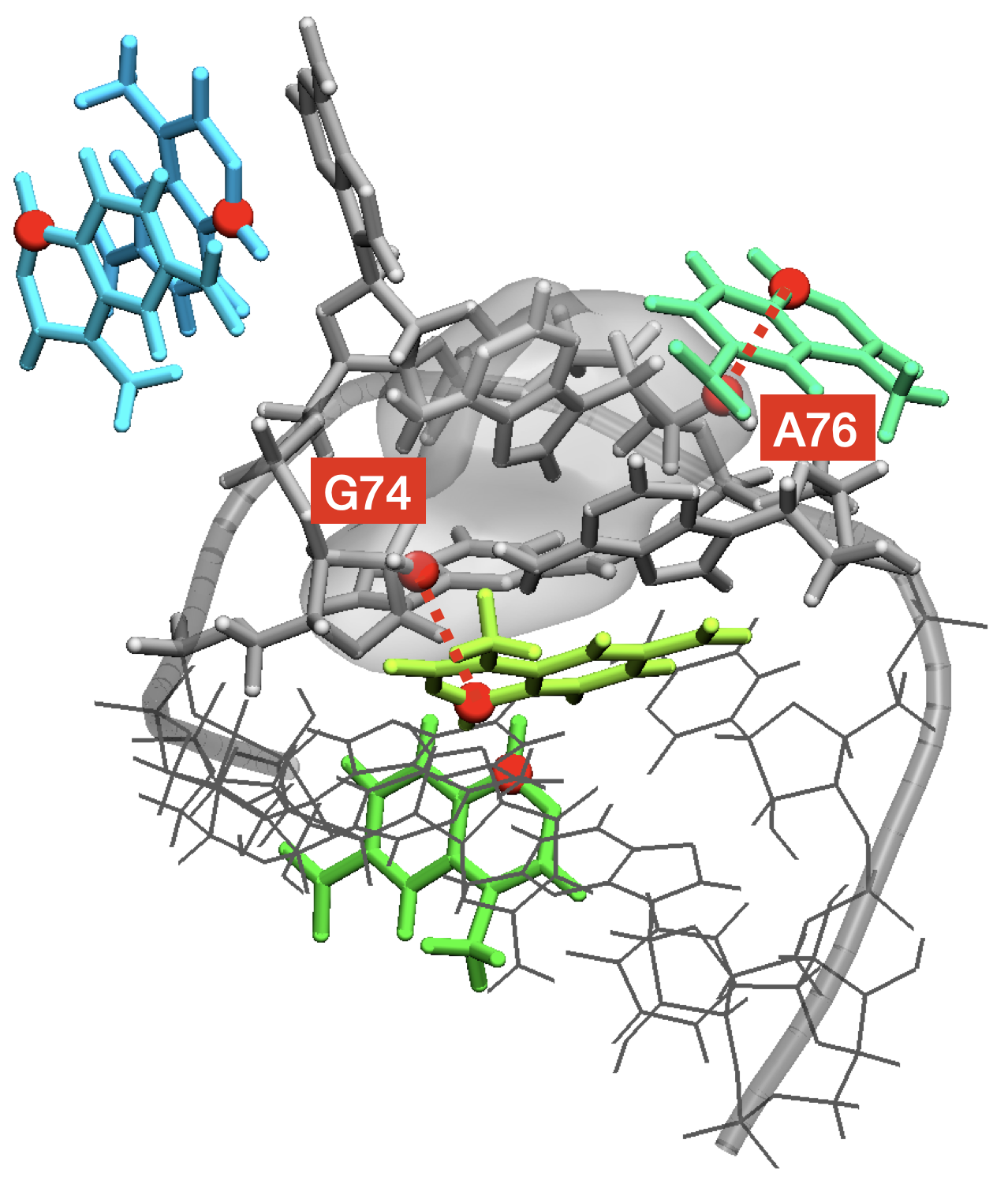}
    \caption{}
    \end{subfigure}\hfill
\begin{subfigure}[ct]{0.33\textwidth}
        \includegraphics[scale = 0.2]{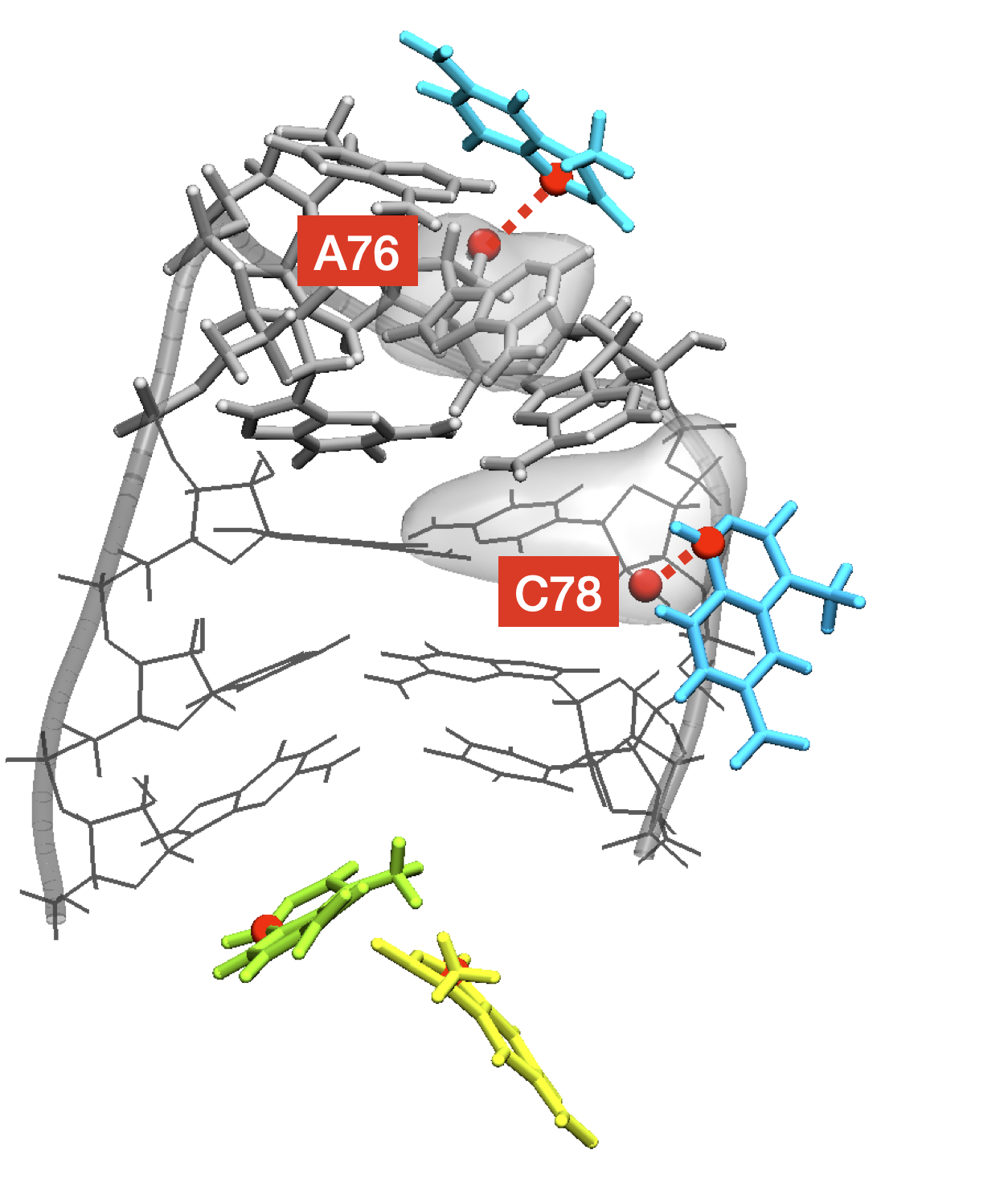}
    \caption{}
    \end{subfigure}
\begin{subfigure}[ct]{0.33\textwidth}
        \includegraphics[scale = 0.2]{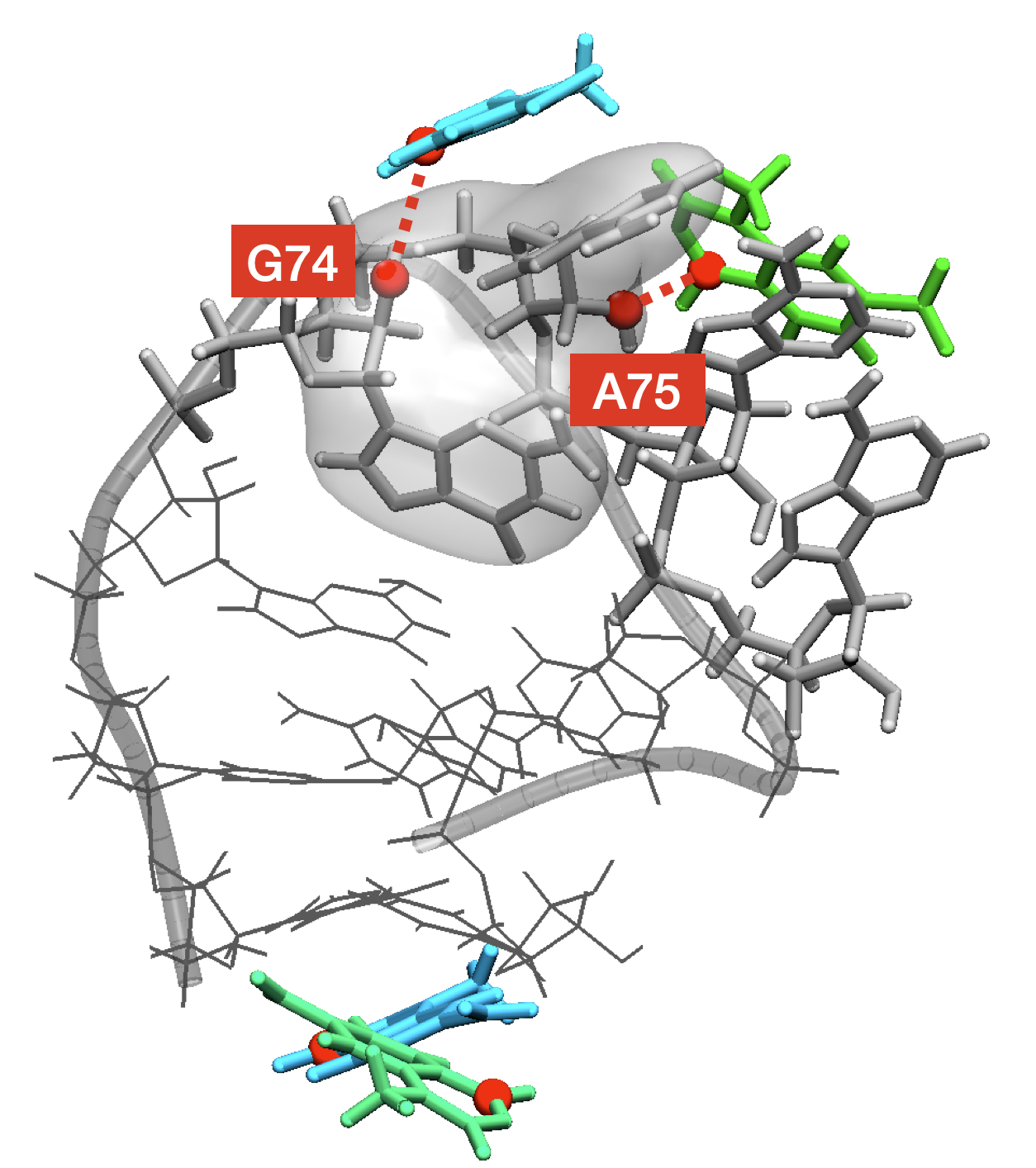}
    \caption{}
    \end{subfigure}
\begin{subfigure}[ct]{0.33\textwidth}
        \includegraphics[scale = 0.2]{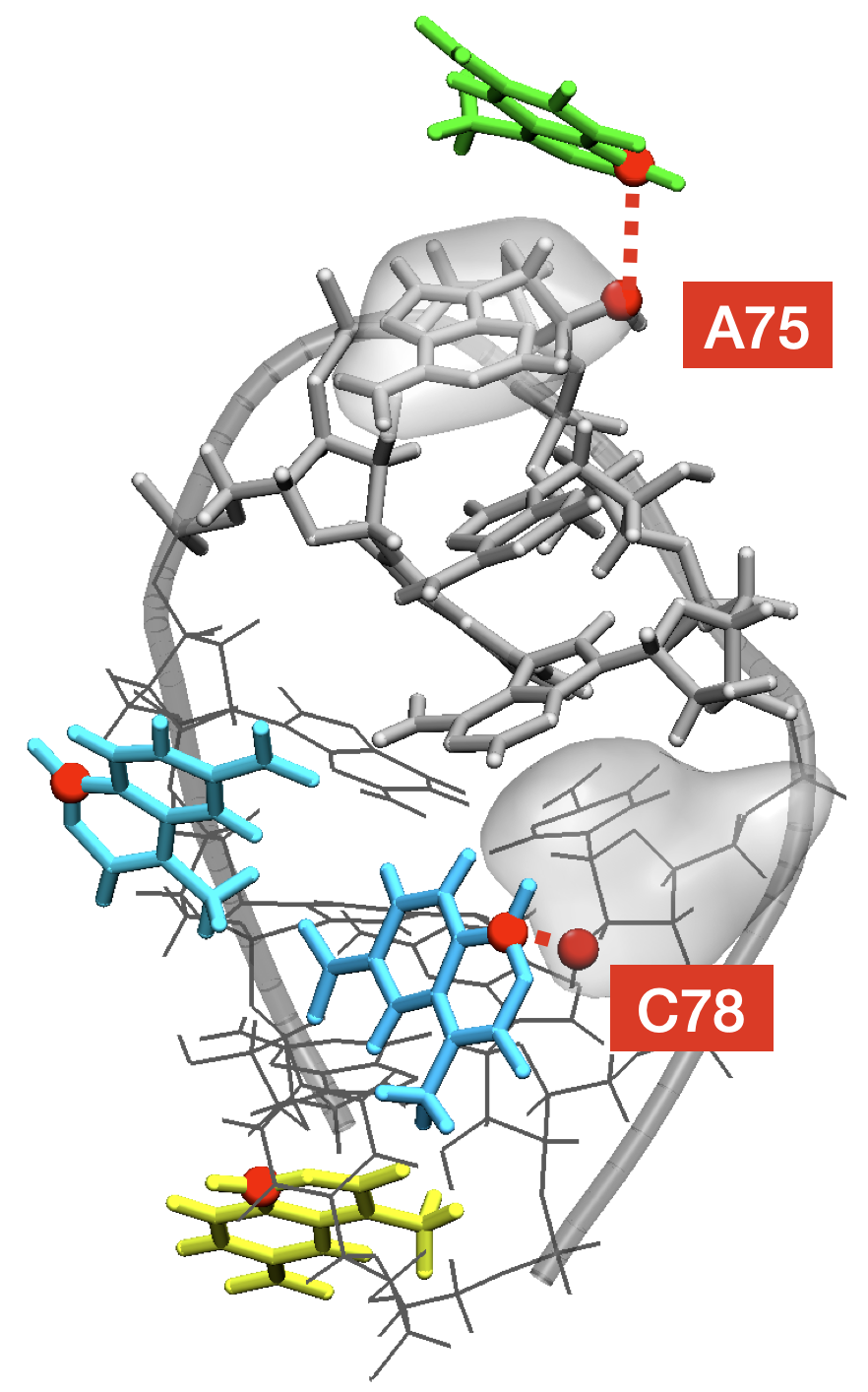}
\caption{}
    \end{subfigure}
\caption{Selection of recurrent conformations of the tetraloop.
         The represented pairs are (a) G73 and G74, (b) G74 and A76, (c) A76 and C78, (d) G74 and A75, (e) A75 and C78,
consistently with Fig.~\ref{dynstru}.
Copies of the reagent are shown in colors, RNA atoms are shown in grey.
Involved nucleotides are shadowed.
}
\label{visualz}
\end{figure*}
In order to investigate the structural signatures of cooperativity, we analyzed the conformations generated in
all the simulated trajectories.
In particular, we extracted sets of frames corresponding to specific conditions and analyzed them using
the \texttt{barnaba} package\cite{bottaro2019barnaba}, which allows showing their dynamic secondary structure representation.
In this representation, base-stacking and base-pairing interactions between nucleotides are reported
accordingly to Leontis-Westhof classification \cite{leontis2001geometric}, and the population of each interaction is shown 
by coloring it accordingly to the reported color map.
Before analysis, frames were subsampled with weights $w(N_A)$ corresponding to their
population at the intermediate concentration  $C=\SI{5.7}{\milli\Molar}$.
Dynamic secondary structures for frames where individual nucleotides were in bound states
are reported in Fig.~\ref{dynstru_single}.
These secondary structures were similar to those obtained by sampling from the
entire trajectory, that are also shown in Fig.~\ref{dynstru_single} as a control,
with limited changes in the populations of the pairings close to the bound site.
Similarly, dynamic secondary structures computed using all frames with weights
that correspond to a range of different values for the reagent concentration
did not show significant effects on RNA structure (see Fig.~\ref{dynstru_conc}),
except for a decrease in the population of the G74:A77 pair.
Then, for each of the cooperative pairs identified in Fig.~\ref{fgr:ddg}, we sampled from the set of frames 
where both nucleotides of the pair where in bound state. Results are shown in
Fig.~\ref{dynstru}.
The only significant change in structure occurs when G74 is involved in a binding event simultaneously
with another cooperative nucleotide in the loop (A75 and A76).
This implies that only when two copies of the reagent are bound to G74 and to one among A75 and A76 the
structural ensemble of the tetraloop is significantly perturbed.
This perturbation is likely responsible for the cooperative effect
reported in Fig.~\ref{fgr:ddg}.
Sample three-dimensional structures corresponding to simultaneous pairs of bound nucleotides are reported in Fig.~\ref{visualz}.

\subsection{Importance of grand-canonical reweighting}

An important advantage of using the grand-canonical reweighting procedure introduced here
with respect to simply consider the finite difference between simulations performed at a different number
of particles is that smooth concentration-dependent curves can be extracted.
In Section \ref{sec:grand-canonical} we compare reactivities obtained at fixed number of copies,
obtained by separately analyzing some of the trajectories discussed above, and
reactivities obtained at fixed chemical potentials, obtained by averaging over the entire
concatenated trajectory. Behavior as a function of the chemical potential is visibly smoother
than behavior as a function of the number of particles, thus making it easier to extract cooperative effects.

\subsection{Control simulations}

We investigated the robustness of our results with respect to two properties
of the simulation settings: helix length and ionic conditions. Since a systematic
study of the effects of these properties on binding cooperativity is
outside the scope of this work, we only generated two control trajectories, 
at a fixed representative value of the number of reagent copies, $N=6$,
and checked that reactivity profiles were consistent with those obtained from
our study.
In one case we simulated a longer part of the SAM-I riboswitch, 
ranging namely from C69 to G82 (two additional base-pairs); in the other case
we increase the ionic strength.
Comparisons with the main simulations are reported respectively
in Section \ref{sec:longer-helix} and \ref{sec:ionic-conditions}.

\subsection{Comparison with experimental analysis}

	\begin{figure}
	  \centering
	  \includegraphics[scale=0.7]{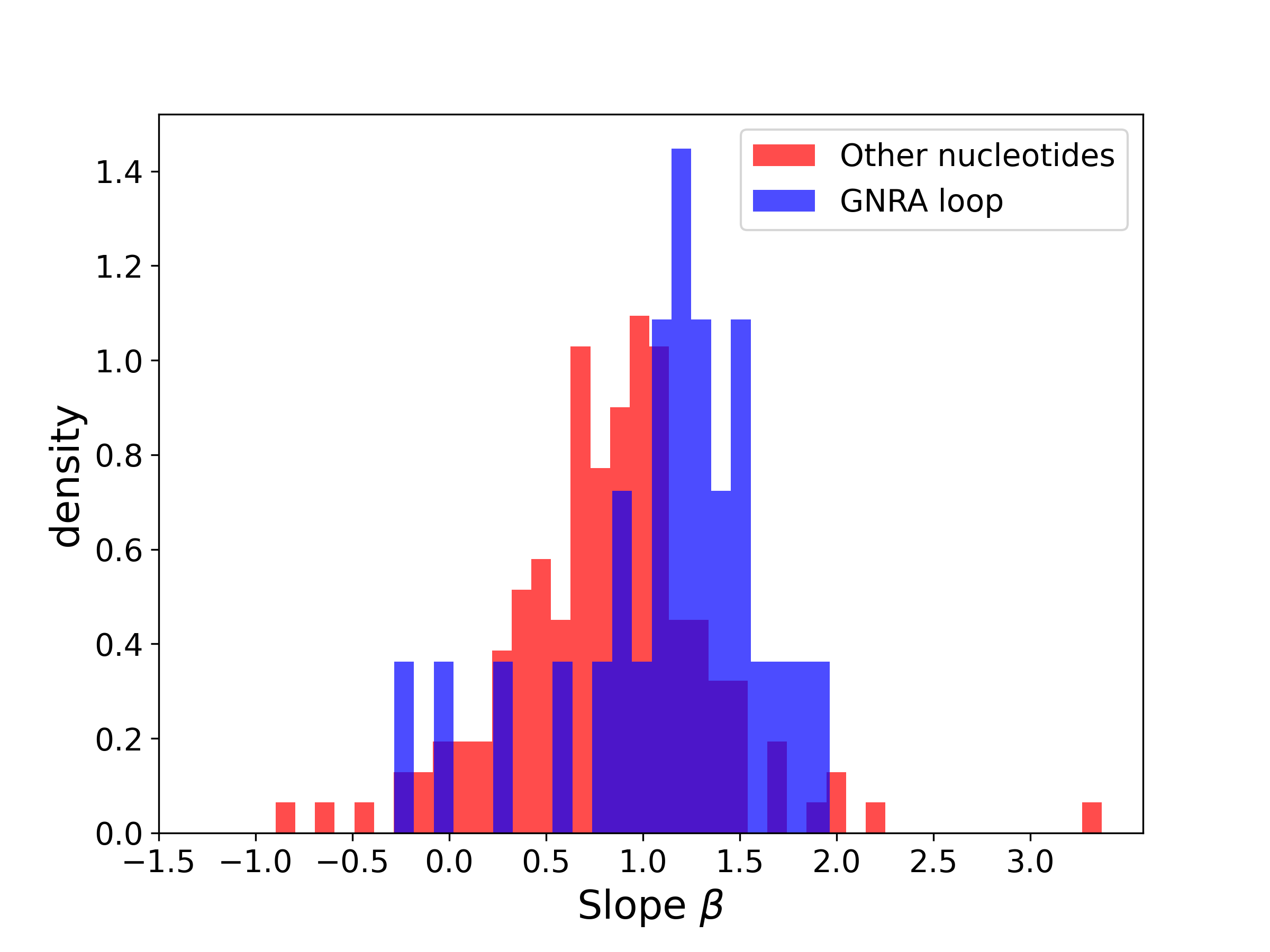}
	  \caption{Distribution of the estimated exponent $\beta$ for concentration-dependent reactivities from experiments.
The coefficient $\beta$ was fitted for each position where the reactivities at the three analyzed concentrations were available.
The blue and red histograms report $\beta$ coefficients obtained for GNRA loops and other nucleotides, respectively.
A coefficient larger than 1 suggests a cooperative effect.
Although results are noisy due to experimental fluctuations and to the fact that the fit was performed using data at only three concentrations,
the distribution for GNRA loops 
		is shifted toward values of $\beta$ greater than one, supporting the cooperative
		reagent binding of these nucleotides observed in the molecular dynamics simulations.}
	  \label{fgr:exp-betas}
	\end{figure}

To validate our hypothesis of cooperative effects in the binding process of SHAPE reagents on RNA,
we analyzed a limited number of experimental datasets.
To the best of our knowledge, chemical probing databases only report
results at one concentration (see, e.g., Ref.~\citenum{cordero2012rna}). We thus generated our own dataset.
This analysis is limited to a small number of RNA structures and reagent concentrations, but suggests that
the effect is experimentally detectable.
We considered a set of SHAPE experiments performed at three difference reagent concentration
(32, 65 and $\SI{125}{\milli\Molar}$) for a set of 5 molecules
for which reference crystallographic or NMR structures are available and can be used
to identify the position of GNRA tetraloops
(2GIS\cite{montange2006structure}, 3GAAA loops,
 1KXK\cite{zhang2002structural}, 1 GAAA loop,
 1SCL\cite{szewczak1993conformation}, 1 GAGA loop,
 1CQ5\cite{schmitz1999structure}, 1 GGAA loop, and
 2GV4\cite{zoll2007breaking}, 1GAAA loop).
We then performed an analysis equivalent to the one reported above
to obtain an exponent $\beta$ associated to each nucleotide.
The coefficients for nucleotides located in GNRA loops are systematically larger than the average (Fig.~\ref{fgr:exp-betas})
supporting the fact that these nucleotides might be affected by cooperative binding effects.

\section{Discussion}

In this work, we use molecular dynamics simulations to identify possible cooperative mechanisms in the binding process of SHAPE
reagents on RNA. We first develop a method to obtain concentration-dependent averages in the grand-canonical ensemble
by combining simulations done with a different number of copies of the reagent. We show how the method works
on a lattice model. Finally, we use it to analyze simulations of 1M7 reagents interacting with a typical RNA structural motif.

The introduced method is based on an idea similar to the one used in weighted-histogram analsysis \cite{ferrenberg1989optimized,kumar1992weighted}, where
a maximum likelihood procedure is used to combine statistics obtained simulating a different number of copies
of the reagent molecule.
We derived and tested our analysis protocol so as to control the chemical potential of a single
molecular species. However, the formalism could be easily extended to multiple molecular species, at the price of setting up a multi-dimensional grid of simulations where the number of copies of each species is scanned.
With respect to the straightforward comparison of simulations performed at different number of particles,
our method has the advantage that it allows to compute properties as smooth functions of the chemical potential and,
thus, of the particle concentration.

At variance with methods based on grand-canonical Monte Carlo \cite{lakkaraju2014sampling,melling2022enhanced} or position-dependent potentials \cite{wang2013grand,perego2015molecular},
the introduced procedure only requires analyzing plain MD simulations. This means that any MD code could be used, and that there will be no overhead associated to changing on-the-fly the number of copies of each molecule in the simulation box or to compute thermodynamic forces to control the number of copies in a given region. However, this advantage comes at a price.
If the actual concentration of the species is unknown, it might be difficult to set up an appropriate range
for the number of copies of molecule to be included in each simulation. Since results will only
be reliable for concentrations that have been actually sampled, this might lead to the need to perform further simulations with a different number of copies. However, also in this case, all simulations could be easily combined to maximize the statistical efficiency.
Similarly to methods based on position-dependent potentials \cite{wang2013grand,perego2015molecular}, our approach is
based on the approximation that subregions of the simulation cell are sufficiently decoupled.
This might limit its applicability in cases where interactions are long ranged, such as electrolyte solutions,
unless ion concentration are large enough to provide a significant screening and make interactions effectively short ranged \cite{finney2021electrochemistry}.
At variance with methods based on position-dependent potentials, however, our procedure allows the region to be selected in the analysis phase.
This in principle allows to fine tune its definition \emph{a posteriori}, without the need to repeat the MD simulations.
An advantage of grand-canonical Monte Carlo methods is that they can be used to effectively enhance the conformational sampling of the controlled species,
which could appear on both sides of a high free-energy barrier. For instance, this would increase sampling of hidden binding pockets.
On the other hand, our method, and methods based on position-dependent potentials, should be explicitly combined with enhanced sampling methods
to cross large free-energy barriers \cite{henin2022enhanced}.
Although in this work we only analyzed plain MD simulations,
enhanced sampling simulations could be analyzed by considering the bias potential when computing the weighting factors.

The method was then applied to the characterization of the dynamics of an RNA structural motif interacting with SHAPE reagents at various concentrations.
Results were obtained by combining 19 independent simulations with a different number of copies of the reagent.
Statistical uncertainties were estimated using Bayesian bootstrapping over the 19 independent simulations.
Since the simulations were prepared independently of each other, the trajectories can be considered as statistically
independent.
This procedure allowed us to define confidence intervals for all the examined quantities.
We were then able to estimate the concentration-dependent probability with which each nucleotide can be bound to a SHAPE reagent.
We also estimated cooperativity effects by analyzing all pairs of nucleotides, showing that pairs of positions located in the RNA tetraloop display
a stronger cooperativity. This cooperativity can be explained as a combination of multiple factors, including interaction between copies of the reagent
and induced changes in the RNA conformational ensemble.
It is worth noting that we didn't observe any pair of positions with a statistically significant anti-cooperativity.
Based on our structural analysis, which shows that reagent binding leads to local RNA destabilization, this is expected.
Inter-reagent stacking can also be reasonably expected to lead to cooperative rather than anti-cooperative effects.
However, we cannot rule out that anti-cooperative effects might arise in more complex structural motifs where,
for instance, physical binding in a position might result in a steric hindrance for binding in a neighboring position,
or even to larger conformational changes of the probed RNA.

The observed cooperativity could be directly detected experimentally, by measuring non-linearities in the dependence of the SHAPE reactivity on
the reagent concentration. Experimental data collected for this work shows that a systematic effect can be observed
where nucleotides located in GNRA tetraloops display a non-linear dependence when compared to the average reactivity,
in qualitative agreement with the results of our simulations.
Whereas the analyzed set of experimental data is limited, this observation suggests that the effect might be general and could be tested
with more systematic experiments performed on a range of reagent concentrations.
A technical but important issue that was not considered here is the fact that
SHAPE reagents are being inactivated by water, resulting in an effective concentration of active reagents that might be lower than the nominal one.
We are not aware of experimental estimates of the concentration of active reagents, and thus we qualitatively used the nominal concentration as a proxy of the effective one.
We additionally note that inactivated (hydrolyzed) reagents are negatively charged, and thus are expected to be
electrostatically repelled by RNA and
less effective than active ones in the crowding effect that is investigated in this work.

An important outcome of this work is that it suggests that different structural motifs might have a different degree of cooperativity.
In this sense, more information could be profitably extracted from experiments performed at different reagent concentrations.
Many different approaches have been suggested to analyze SHAPE reactivities to improve RNA structure prediction,
including the idea of identifying reactivity patterns for known motifs \cite{cao2021characteristic}
and of combining data obtained with different reagents\cite{rice2014rna,saaidi2020ipanemap}.
However, we are not aware of any attempt to use concentration-dependent information as it is suggested here.
The measurement of concentration-dependent reactivities for a sufficently
large number of training RNA systems of known structure is left as a subject for a future work.

\begin{acknowledgement}

David H.~Mathews and Karissa Sanbonmatsu are acknowledged for reviewing a very preliminary version of this work,
which is available at \url{http://hdl.handle.net/20.500.11767/116273},
and providing several suggestions.

\end{acknowledgement}

\begin{suppinfo}
Derivation of the algorithm with likelihood maximization;
relationship between chemical potential and concentration;
results on a lattice model;
parameters for 1M7;
sample initial conformations;
structural analysis at various concentrations;
effects of grand-canonical reweighting, ionic conditions, and system size.

\clearpage

\renewcommand\thealgorithm{S\arabic{algorithm}}    
\renewcommand\thefigure{S\arabic{figure}}    
\setcounter{figure}{0} 
\renewcommand\thetable{S\arabic{table}}    
\setcounter{table}{0} 
\renewcommand\thesection{S\arabic{section}}    
\setcounter{section}{0} 

	\section{Likelihood maximization}
        \label{sec:likelihood-maximization}
	The likelihood of observing $\mathbf{t}=\{t_{Nk}\}$ frames from a set 
	$\mathcal{S}$ of $N_{max}$ independent simulations, with $t_{Nk}$ frames with $k$ particles 
	in region $A$ and $N-k$ particles in region $B$, is
	\begin{equation}
		P\left(\mathbf{t}|\Omega_A,\Omega_B,\{c_N\}\right)\propto\prod\limits_{N=1}^{N_{max}}\prod\limits_{k=0}^{N}\left(c_N\Omega_A(k)\Omega_B(N-k)\right)^{t_{Nk}}
	\end{equation}
	up to a normalization constant. The normalization coefficients $\{c_N\}$ are required 
	to ensure that, for each value of the number of reagent copies $N$
	\begin{equation}
		\sum\limits_{k=0}^N c_N\Omega_A(k)\Omega_B(N-k)=1
	\end{equation}
	Taking the negative logarithm of the likelihood and adding Lagrangian multipliers $\{\lambda_N\}$
	to ensure the above normalization requirement yields
	\begin{multline}
		\mathcal{L}\left(\mathbf{t}|\Omega_A,\Omega_B,\{c_N\}\right)= \\
		-\sum\limits_{N=1}^{N_{max}}\sum\limits_{k=0}^{N} t_{Nk}\log\left(c_N\Omega_A(k)\Omega_B(N-k)\right)
		-\sum_N \lambda_N\left(\sum\limits_k c_N\Omega_A(k)\Omega_B(N-k)-1\right)
	\end{multline}
	The Lagrangian function $\mathcal{L}$ can be minimized by setting to zero its gradient 
	with respect to the model parameters $\Omega_A$, $\Omega_B$, $\{c_N\}$, and to the Lagrangian
	multipliers $\{\lambda_N\}$. The resulting equations are:
	\begin{subequations}
		\begin{align}
			&\frac{\partial\mathcal{L}}{\partial \lambda_N}=\sum\limits_k c_N\Omega_A(k)\Omega_B(N-k)-1=0\label{dln}\\
			&\frac{\partial\mathcal{L}}{\partial c_N}=-\frac{\sum_k t_{Nk}}{c_N}-\lambda_N\sum_k \Omega_A(k)\Omega_B(N-k)=0\label{dcn}\\
			&\frac{\partial\mathcal{L}}{\partial \Omega_A(k)}=-\frac{\sum_N t_{Nk}}{\Omega_A(k)}-\sum_N\lambda_N c_N \Omega_B(N-k)=0\label{doa}\\
		&\frac{\partial\mathcal{L}}{\partial \Omega_B(k)}=-\frac{\sum_N t_{N,N-k}}{\Omega_B(k)}-\sum_N\lambda_N c_N \Omega_A(N-k)=0\label{dob}
		\end{align}
	\end{subequations}
	From the equation \ref{dln}, we obtain
	\begin{equation}
		\sum_k\Omega_A(k)\Omega_B(N-k)=\frac{1}{c_N}
	\end{equation}
	that we replace in the second term of the equation \ref{dcn} yielding
	\begin{equation}
                \label{eq:lambda}
		\lambda_N=-\sum_k t_{Nk}
	\end{equation}
	We then define: $A_k=\sum_Nt_{Nk}$, counting the number of times that,
	in the whole set of $N_{max}$ trajectories, a particle was found in region $A$;
        $B_k=\sum_Nt_{N,N-k}$, counting the equivalent number for region $B$;
	and $L_N=\sum_k t_{Nk}$, the total number of frames accumulated in the trajectory with $N$ particles.
        By substituting Eq.~\ref{eq:lambda} and the definitions of $A_k$, $B_k$, and $L_N$ in Equations \ref{doa} and \ref{dob} we obtain
	\begin{equation}\label{omegaequations}
		\begin{aligned}
			&\Omega_A(k)=\frac{A_k}{\sum_N L_N c_N \Omega_B(N-k)}\\
			&\Omega_B(k)=\frac{B_k}{\sum_N L_N c_N \Omega_A(N-k)}
		\end{aligned}
	\end{equation}
	These equations can be solved iteratively through the algorithm reported in Alg.~\ref{gcsolve}.
	\begin{algorithm}
		\caption{Estimating $\Omega_A$ and $\Omega_B$}\label{gcsolve}
		\begin{algorithmic}[1]
			\State $\Omega^{i=0}_A\left[k\right]\gets A\left[k\right] \quad \forall k\in\Big[0,\dots,N_{max}\Big]$
			\State $\Omega^{i=0}_B\left[k\right]\gets B\left[k\right] \quad \forall k\in\Big[0,\dots,N_{max}\Big]$
			\State $\text{threshold}\gets 10^{-30}$
			\For{$i\in\{1,\dots, N_{steps}\}$}
		  	\State $c[N]\gets 1/\sum_{k=0}^{N}\Omega^{(i-1)}_A\left[k\right]\cdot\Omega^{(i-1)}_B\left[N-k\right]\quad \forall N\in \Big[1,\dots,N_{max}\Big]$
		  	\State $\Omega^{(i)}_A\left[k\right]\gets A\left[k\right]/\sum_{N=k}^{N_{max}}L[N]\cdot c[N]\cdot\Omega^{(i-1)}_B\left[N-k\right]
		  	\quad \forall k\in\Big[0,\dots,N_{max}\Big]$
		  	\State $\Omega^{(i)}_B\left[k\right]\gets B\left[k\right]/\sum_{N=k}^{N_{max}}L[N]\cdot c[N]\cdot\Omega^{(i-1)}_A\left[N-k\right]
		  	\quad \forall k\in\Big[0,\dots,N_{max}\Big]$
		  	\State $\varepsilon \gets \sum_{k=0}^{N_{max}}\Big[\left(\Omega^{(i)}_A[k]-\Omega^{(i-1)}_A[k]\right)^2+\left(\Omega^{(i)}_B[k]-\Omega^{(i-1)}_B[k]\right)^2\Big]$
		  	\State $\Omega^{(i)}_B\gets \Omega^{(i)}_B/\Omega^{(i)}_B[0]$\label{norm1}
		  	\State $f\gets \Omega^{(i)}_B[1]/\Omega^{(i)}_B[0]$
		  	\State $\Omega^{(i)}_B[k]\gets \Omega^{(i)}_B[k]/(f^k B[0]) \quad \forall k\in\Big[0,\dots,N_{max}\Big]$
		  	\State $\Omega^{(i)}_A[k]\gets \Omega^{(i)}_A[k]/(f^k A[0]) \quad \forall k\in\Big[0,\dots,N_{max}\Big]$\label{norm2}
		  	\If{$\varepsilon<\text{threshold}$}
		  	\State\textbf{break}
		  	\EndIf
			\EndFor
		\end{algorithmic}
	\end{algorithm}
	Noticeably, line \ref{norm1} to \ref{norm2} provide a normalization of $\Omega_A$ and $\Omega_B$ such that
	$\Omega_A\left(k=0\right)=1$ and $\Omega_B\left(k=0\right)=\Omega_B\left(k=1\right)=1$. In this way the free energy of the state
	with no particle at all is set to zero, as well as the free-energy cost for adding the first particle to region $B$. Since
	the chemical potential $\mu$ is defined up to a constant, $\Omega_A$ and $\Omega_B$ are invariant with respect to scaling by an
	arbitrary factor $f$ each, and to scaling each $k$-th component of $\Omega_{A/B}$ by the $k$-th power of the same factor $f^k$.
	By choosing $f=\Omega_B\left(k=1\right)/\Omega_B\left(k=0\right)$ the normalization is fixed and the scaling invariance is removed. This scaling has no impact on the final weights, but can affect the relationship between the chemical potential and the particle concentrations
and can be used to assign a physical interpretation to the obtained canonical partition functions.
\clearpage
	\section{Fixing reagent concentration through the chemical potential}
        \label{sec:fixing-mu-of-n}
	Once the ML estimates of $\Omega_A$ and $\Omega_B$ are computed, estimates of grand-canonical averages
	of any function $f(k)$ of the number of particles $k$ in either region $A$ or $B$ can be computed as 
	\begin{equation}
		\langle f(k)\rangle_{GC}=\sum\limits^{N_{max}}_{k=0}f(k)\cdot P^{GC}_{A/B}\left(k\right)
		   =\frac{\sum\limits^{N_{max}}_{k=0}f(k)\cdot\Omega_{A/B}(k)e^{-\mu k/RT}}{\sum\limits_{k=0}^{N_{max}}\Omega_{A/B}(k)e^{-\mu k/RT}}
	\end{equation}
	where $\mu$ is the chemical potential. 
	It is thus straightforward to compute the grand-canonical average of the number of reagent copies
	in both regions by replacing $f(k)=k$ in the equation above. 
	Since we want to compute averages of quantites such as affinity and cooperativity at varying reagent
	concentration, we first have to identify the values of $\mu$ corresponding to the desired concentrations, using 
	the algorithm reported in Alg.~\ref{musolve}.
	\begin{algorithm}
		\caption{Estimating $\mu$ as function of the number of particles in $B$}\label{musolve}
		\begin{algorithmic}[1]
			\Function{nb\_of\_mu}{$\mu$}
			\State $OB\gets\Omega_B$\text{ obtained using Alg.~\ref{gcsolve}}
			\State $NB\gets\text{desired }N_B$
			\State $P_B[k]\gets OB[k]\cdot e^{-\mu k/RT}\quad\forall k\in\Big[0,\dots,N_{max}\Big]$
			\State $P_B\gets P_B/\sum\limits_{k=0}^{N_{max}}P_B$
			\State $N_B^{est}\gets\sum\limits_{k=0}^{N_{max}}k\cdot P_B[k]$\\
			\State\Return $\log{N_B^{est}}-\log{NB}$
			\EndFunction
			\State find the root of \textproc{nb\_of\_mu} through an optimized bisection routine
		\end{algorithmic}
	\end{algorithm}
\clearpage
	\section{Lattice model}
        \label{sec:ci:lattice}
	The introduced methodology for grand-canonical reweighting of molecular dynamics presented in the Methods section in the main text 
	is first tested on a lattice model. We consider a lattice space
	divided in two regions, $A$ and $B$. Region $A$ contains $S_A$ sites and region $B$ contains $S_B$ sites. Sites are then populated
	with a varying number of particles that interact with each other only through mutual exclusion, so that a site cannot be
	occupied by more than one particle.
        We first consider two possible scenarios:
 a purely entropic systems, in which all the sites are equivalent and the free energy
	depends only on the entropic contribution of the number of possible combinations of up to $N_{max}$ particles occupying the
	$S=S_A+S_B$ sites; and a system in which the presence of a stabilizing site in the lattice region $A$ brings in an additional
	energetic contribution to the free energy.
	In both cases, the partition functions $\Omega_A$ of region $A$ and $\Omega_B$ of region $B$ can be computed analitically. These functions are
	normalized as explained in Section \ref{sec:likelihood-maximization}, that is by setting $\Omega_A(k=0)=\Omega_B(k=0)=1$, and $\Omega_B(k=1)=1$, so that
	the zero of free energy corresponds to the empty lattice and the free-energy cost for insertion of the first particle in region
	B is set to zero. The normalization is accomplished by scaling each $\Omega_{A/B}(k)$ by a factor $1/(f^k\Omega_{A/B}(0))$, where $f=\left(\frac{\Omega_B(1)}{\Omega_B(0)}\right)$.
	\par
	In the purely entropic system, the two partition functions are related to the number of different combinations
	in which particles can be distributed in the sites:
	\begin{equation}
		\Omega_{A/B}(k)\propto\binom{S_{A/B}}{k}=\frac{S_{A/B}!}{(S_{A/B}-k)!k!}
	\end{equation}
	If the number of sites in $A$ and $B$ is equal, then populating a site in $A$ has the same free-energy cost of populating one in $B$.
	\begin{figure*}
		\centering
		\begin{subfigure}{0.49\linewidth}
			\includegraphics[width=1.0\textwidth, keepaspectratio]{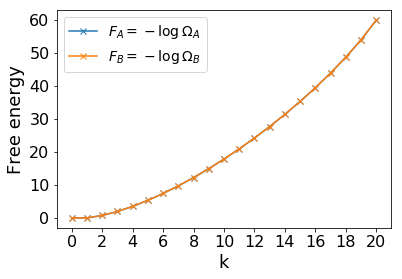}
			\caption{}
			\end{subfigure}\hfill
		\begin{subfigure}{0.49\linewidth}
			\includegraphics[width=1.0\textwidth, keepaspectratio]{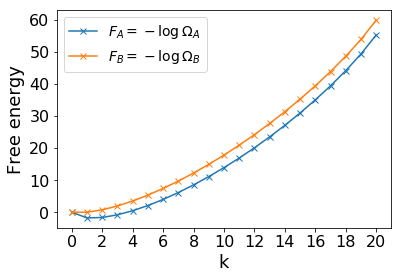}
			\caption{}
		\end{subfigure}
		\caption{Free-energy contributions of the two regions $A$ and $B$ of a lattice space populated with mutually exclusive
	         particles. In a purely entropic system with $A$ as large as $B$ (panel~a), the two lines are indistinguishable.
                 In a system with
	         a stabilizing site in region $A$ (panel~b), the free energy gain associate to insertion in region $A$ rather than region $B$ can be seen.}
	\label{lattice_ene}
	\end{figure*}
	The free-energy contributions of the two regions $F_{A/B}=-\log\Omega_{A/B}$  for numbers of sites $S_A=S_B=20$ is shown in Fig.~\ref{lattice_ene}a.
	\par
	In the second scenario, regions $A$ and $B$ differ for the presence of a single stabilizing site in $A$, for which the probability 
	to be populated is 100 times larger than the other sites, which corresponds to a stabilization of $-RT\log100$.
	\begin{equation}
		\Omega_{A}(k)\propto\binom{S_{A}-1}{k}+100 \binom{S_{A}-1}{k-1}
	\end{equation}
	As shown in Fig.~\ref{lattice_ene}b,
        when at least one particle is present in the lattice,
	the presence of the stabilizing site systematically contributes with a
	free-energy gain ($-\log \Omega_A(k) < -\log \Omega_B(k)$).
	Since the second case is more representative of a molecular system where particles can interact with a
a solute molecule in region $A$, we consider only this case for the following tests, but similar results could be obtained for the purely entropic system.

We first address the limitations of the method arising from the fact that a finite $N_{max}$ is used.
We assume to be able to collect the information about the probability of observing a given number of particles in region $A$ or $B$
from a set of simulations performed with a fixed number of particles ranging from 1 to $N_{max}$.
These probabilities are
$A_k\propto \sum_{N=1}^{N_{max}} \frac{\Omega_A(k)\Omega_B(N-k)}{\sum_{k'}\Omega_A(k')\Omega_B(N-k')}$ and
$B_k\propto \sum_{N=1}^{N_{max}} \frac{\Omega_A(N-k)\Omega_B(k)}{\sum_{k'}\Omega_A(k')\Omega_B(N-k')}$ 
for regions $A$ and $B$, respectively.
These probabilities, which correspond to the histogram that one could accumulate in a set of infinitely long (perfect sampling) simulations at fixed number of particles,
are then used to infer estimates for $\Omega_A$ and $\Omega_B$ using Algorithm \ref{gcsolve}.
Figure~\ref{latticeomega} shows the exact and inferred $\Omega$ for the two regions.
The inference is exact when $k\leq N_{max}$. However, since we are in a regime where none of the analyzed simulations has more than $N_{max}$ particles,
the method has no way to infer the partition function of $k>N_{max}$.
	\par
	\begin{figure*}
		\centering
		\begin{subfigure}{0.49\linewidth}
		\includegraphics[width=1.0\textwidth, keepaspectratio]{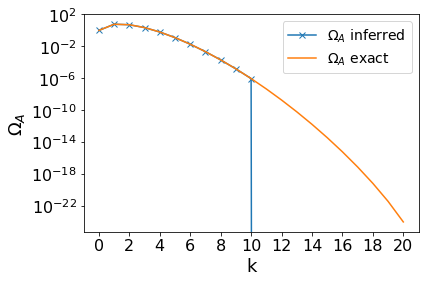}
		\caption{}
		\end{subfigure}\hfill
		\begin{subfigure}{0.49\linewidth}
		\includegraphics[width=1.0\textwidth, keepaspectratio]{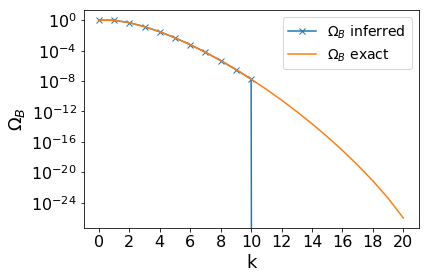}
		\caption{}
		\end{subfigure}
		\caption{Inferred and exact values of the partition functions (a) $\Omega_A$ of region A
		        and (b) $\Omega_B$ of region $B$, with infinite sampling and number of particles $N_{max}=10$ lower than
		        the number of lattice sites, $S=40$. Whereas for $k\leq N_{max}$ the inference is exact,
		        the inferred $\Omega_{A/B}(k)=0 \;\forall k>N_{max}$.}
		\label{latticeomega}
	\end{figure*}

We then use the inferred partition functions to compute the distribution of particles in region $A$ at fixed concentration of particles in region $B$.
This requires to first estimate the chemical potential $\mu$ associated to the concentration in region $B$ using Algorithm \ref{musolve}, and then
computing the probability to observe $k$ particles in region $A$ using the grand-canonical weights at that chemical potential.
Figure \ref{nmaxeffects}a shows the estimated average number of particles in each region at fixed chemical potential.
The inferred number of particles agrees very well with the analytical solution for values of $\mu$ corresponding to a number of particles
smaller than $N_{max}=10$. However, at a chemical potential low enough to lead to more than 10 particles per region, the inference is incorrect
and predicts only 10 particles per region at most. The inferred partition functions are then used to compute the distribution
of the number of particles in region $A$ for a list of 5 different concentrations, which correspond to a growing average number of particles
in region $B$ (Figure \ref{nmaxeffects}, panels b--f).
When the concentration corresponds to an average number of particles in $B$ significantly smaller than $N_{max}=10$ (panels b--d), the agreement between the inferred and 
exact solution is virtually perfect. The limitations of the method are clear in cases where the exact distribution would imply
a non neglibigle probability to observe more than 10 particles in region $A$ (panel e). Panel f represents the extreme case, where the inferred chemical potential
is $-\infty$ and the inferred distribution only allows 10 particles in region $A$. This test highlights that results can only be reweighted to concentrations that are compatible with
the number of particles included in the analyzed simulations.

	\begin{figure*}
	\centering
	\begin{subfigure}{0.33\linewidth}
	\includegraphics[width=1.0\textwidth, keepaspectratio]{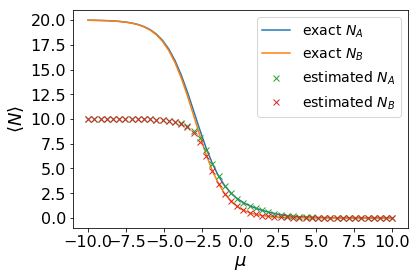}
	\caption{}
	\end{subfigure}\hfill
	\begin{subfigure}{0.33\linewidth}
	\includegraphics[width=1.0\textwidth, keepaspectratio]{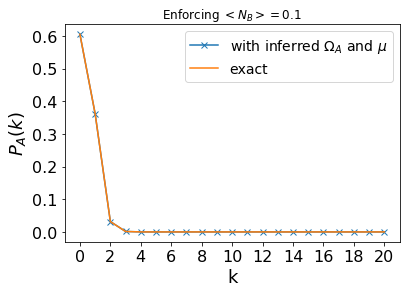}
	\caption{}
	\end{subfigure}\hfill
	\begin{subfigure}{0.33\linewidth}
	\includegraphics[width=1.0\textwidth, keepaspectratio]{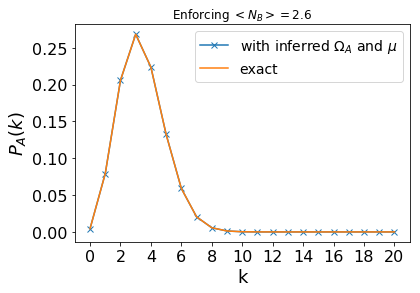}
	\caption{}
	\end{subfigure}\vfill
	\begin{subfigure}{0.33\linewidth}
	\includegraphics[width=1.0\textwidth, keepaspectratio]{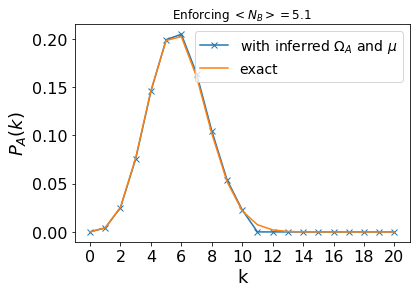}
	\caption{}
	\end{subfigure}\hfill
	\begin{subfigure}{0.33\linewidth}
	\includegraphics[width=1.0\textwidth, keepaspectratio]{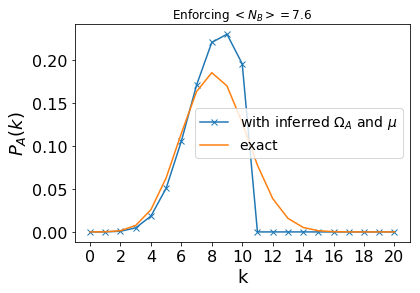}
	\caption{}
	\end{subfigure}\hfill
	\begin{subfigure}{0.33\linewidth}
	\includegraphics[width=1.0\textwidth, keepaspectratio]{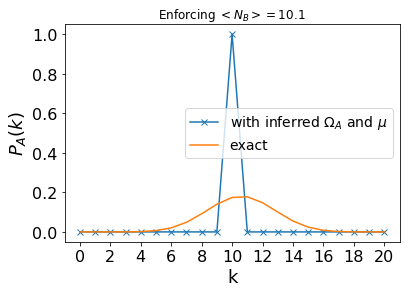}
	\caption{}
	\end{subfigure}
	\caption{Limitations in the estimates of (a) chemical potential at desired concentrations $\mu(N_B)$
	         in region $B$, and (b-f) of the probability distribution of the number of particles in region $A$ at
	         different values of the enforced concentration in $B$, as tested on the lattice model with a
	         stabilizing site in region $A$.}
	\label{nmaxeffects}
	\end{figure*}
	\par
	\begin{figure*}
		\centering
		\begin{subfigure}{0.45\linewidth}
		\includegraphics[width=1.0\textwidth, keepaspectratio]{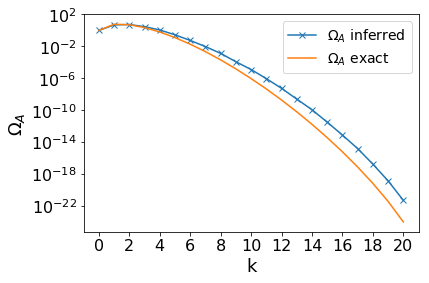}
		\caption{}
		\end{subfigure}\hfill
		\begin{subfigure}{0.45\linewidth}
		\includegraphics[width=1.0\textwidth, keepaspectratio]{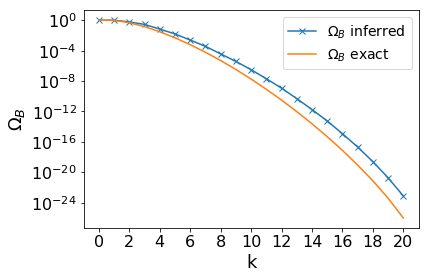}
		\caption{}
		\end{subfigure}\vfill
		\begin{subfigure}{0.33\linewidth}
		\includegraphics[width=1.0\textwidth, keepaspectratio]{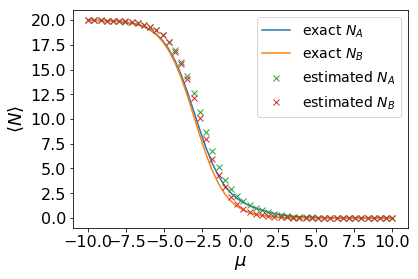}
		\caption{}
		\end{subfigure}\hfill
		\begin{subfigure}{0.33\linewidth}
		\includegraphics[width=1.0\textwidth, keepaspectratio]{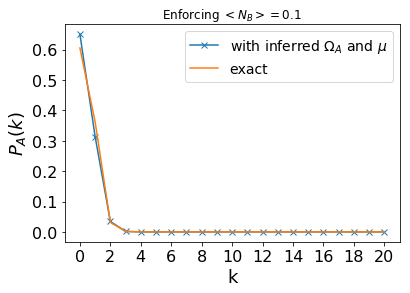}
		\caption{}
		\end{subfigure}\hfill
		\begin{subfigure}{0.33\linewidth}
		\includegraphics[width=1.0\textwidth, keepaspectratio]{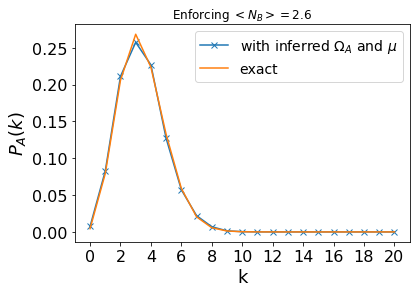}
		\caption{}
		\end{subfigure}\vfill
		\begin{subfigure}{0.33\linewidth}
		\includegraphics[width=1.0\textwidth, keepaspectratio]{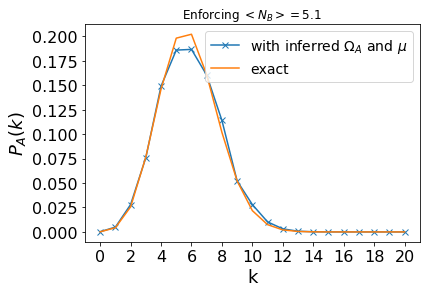}
		\caption{}
		\end{subfigure}\hfill
		\begin{subfigure}{0.33\linewidth}
		\includegraphics[width=1.0\textwidth, keepaspectratio]{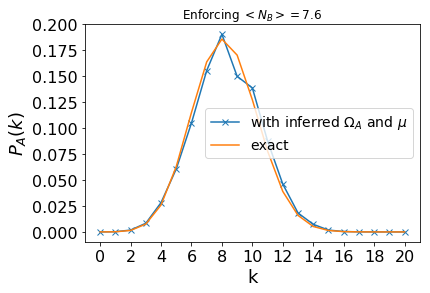}
		\caption{}
		\end{subfigure}\hfill
		\begin{subfigure}{0.33\linewidth}
		\includegraphics[width=1.0\textwidth, keepaspectratio]{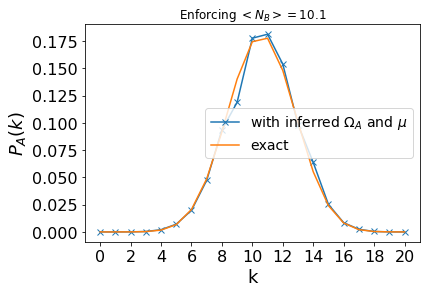}
		\caption{}
		\end{subfigure}
		\caption{Finite sampling effects in the estimations of (a) $\Omega_A(k)$, (b) $\Omega_B(k)$, (c) the relation
		         between chemical potential $\mu$ and number of particles in region $A$ and $B$, and of (d-h) the probability
		         distribution $P_A(k)$ of the number of particles in region $A$, as tested on the lattice model with a
		         stabilizing site in region $A$.}
		\label{finitesampling}
	\end{figure*}

Next, we address the issue of obtaining histograms using a finite number of samples.
To this aim, instead of assuming to have access to the exact $A_k$ and $B_k$, we estimate them by drawing
$L=100$ samples from the exact probability distributions. In other words, we consider $N_{max}$ simulations
accumulating 100 independent samples. To remove the limitation associated to the finite value of $N_{max}$ discussed above,
we here consider $N_{max}=20$.
We first show the inferred partition functions (Figure \ref{finitesampling}a and b).
Since the number of simulated particles is sufficient to cover all cases, no discontinuity is observed.
However, the inferred $\Omega$ does not match anymore the exact reference.
To have an idea of how much this error on the inference of $\Omega$ would affect the final result,
we use the inferred $\Omega$ to estimate the dependence of the average number of particles on $\mu$ (Figure \ref{finitesampling}c).
The impact is minimal now, since a sufficient number of particles have been included. 
The resulting estimates for the population of region $A$ at different particle concentrations are only slightly affected (panels d--h).
Clearly, this effect would be larger if the number of samples per simulation was chosen to be smaller.

\clearpage
	\section{Parametrization of 1M7}
	The partial charges of 1-methyl-7-nitroisatoic anhydride (1M7) computed through RESP
	as described in the main text, are reported in Table~\ref{par1m7} 
	\begin{table}
		\centering
		\begin{tabular}{l l}
			Atom & Charge (e) \\
			\hline
			C1 & 0.013677 \\ 
			H1 & 0.165938 \\ 
			C2 & -0.301276 \\
			H2 & 0.214747 \\
			C3 & 0.213481 \\
			N2 & 0.680265 \\
			O4 & -0.418660 \\
			O5 & -0.418660 \\
			C4 & -0.403042 \\
			H3 & 0.221441 \\
			C5 & 0.431676 \\
			N1 & -0.308580 \\
			C8 & 0.913611 \\
			O2 & -0.562770 \\
			C9 & -0.336386 \\
			H4 & 0.158142 \\
			H5 & 0.158142 \\
			H6 & 0.158142 \\
			C6 & -0.413908 \\
			C7 & 0.925472 \\
			O3 & -0.538891 \\
			O1 & -0.552561 \\
			\hline
		\end{tabular}
		\caption{Charges of atoms in the 1M7 topology as obtained via Antechamber using the RESP method.}
		\label{par1m7}
	\end{table}
\clearpage
	\section{Initial conformations}
	Examples of the initial conformation for simulations with $N=5$ and $N=16$ are reported in 
	Fig.~\ref{start_conf}.
	\begin{figure*}
	\centering
	    \begin{subfigure}[t]{0.45\textwidth}
	        \includegraphics[width=1.05\linewidth, keepaspectratio]{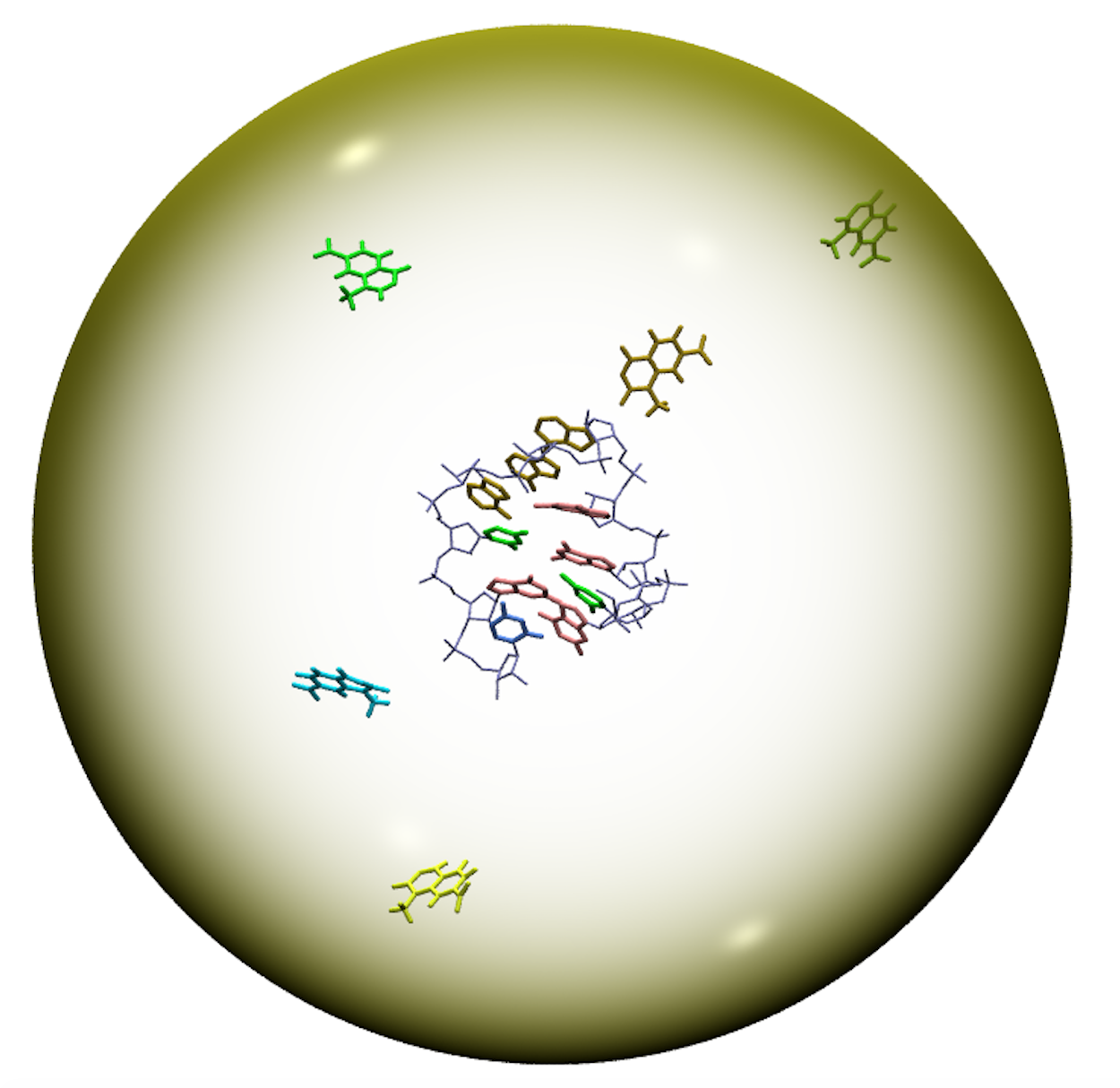}
	        \caption{}
	    \end{subfigure}\hfill
	    \begin{subfigure}[t]{0.45\textwidth}
	        \includegraphics[width=1.15\linewidth, keepaspectratio]{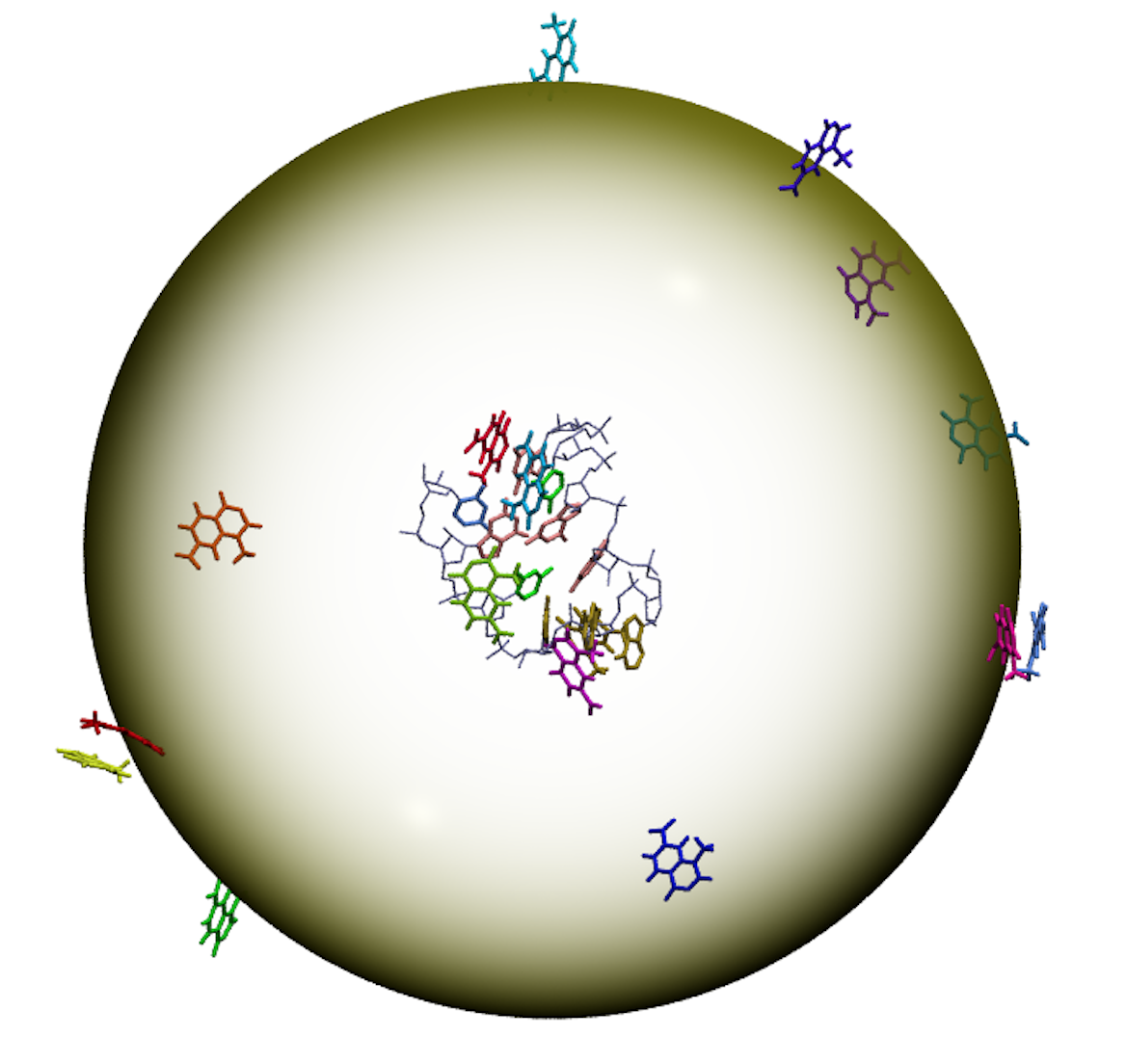}
	        \caption{}
	    \end{subfigure}
	    \caption{Two examples of initial conformation for the simulations of the tetraloop with
	         (a) $N=5$ and (b) $N=16$ probes. The yellow sphere represents the surface around the tetraloop
	         where probes are initially placed. The RNA tetraloop is at center of the sphere and is shown 
		 in thin sticks representation, with nucleobases highlighted as thicker sticks and with different colors. 
		 Different 1M7 molecules are displayed with different colors.}
	    \label{start_conf}
	\end{figure*}
\clearpage
	\section{Structural variability at varying reagent concentration}
We here report an analysis similar to the one reported in Figs.~\ref{dynstru_single} and~\ref{dynstru}.
Here, no assumption is done on reagent binding.

	\begin{figure*}[h!]
	\centering
	\begin{subfigure}[t]{0.33\textwidth}
	        \includegraphics[scale=0.20]{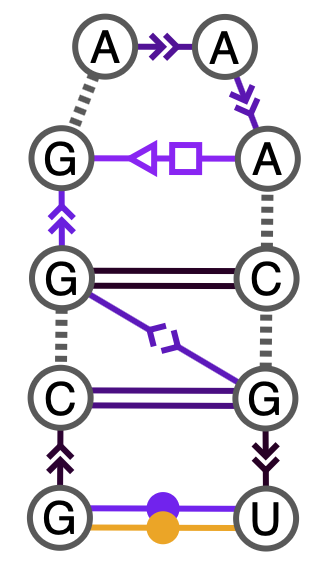}
	    \caption{}
	    \end{subfigure}\hfill
	    \begin{subfigure}[t]{0.33\textwidth}
	        \includegraphics[scale=0.20]{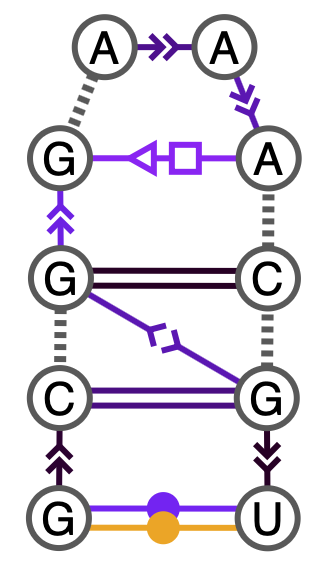}
	\caption{}
	    \end{subfigure}\hfill
	    \begin{subfigure}[t]{0.33\textwidth}
	        \includegraphics[scale=0.20]{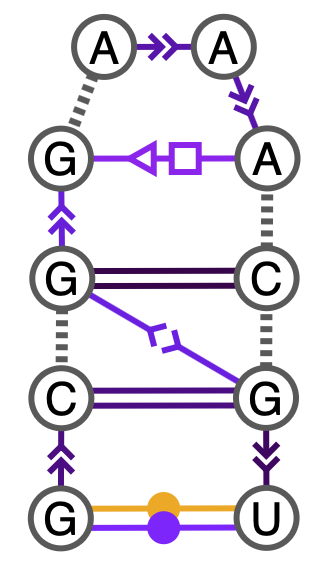}
	    \caption{}
	    \end{subfigure}\vfill
	    \begin{subfigure}[t]{0.33\textwidth}
	        \includegraphics[scale=0.20]{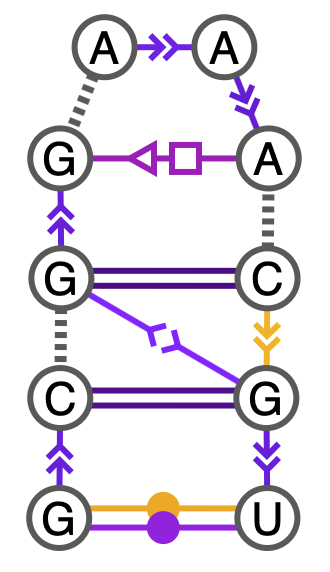}
	\caption{}
	    \end{subfigure}\hfill
	    \begin{subfigure}[t]{0.33\textwidth}
	        \includegraphics[scale=0.20]{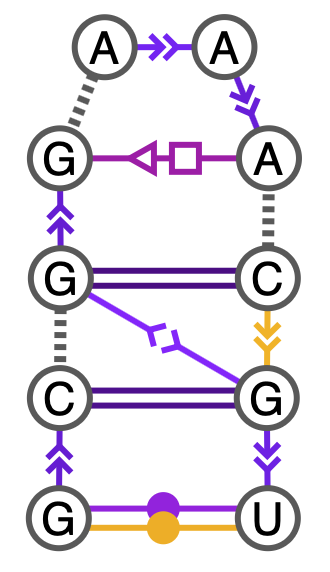}
	\caption{}
	    \end{subfigure}\hfill
	    \begin{subfigure}[t]{0.33\textwidth}
		    \includegraphics[scale=0.20]{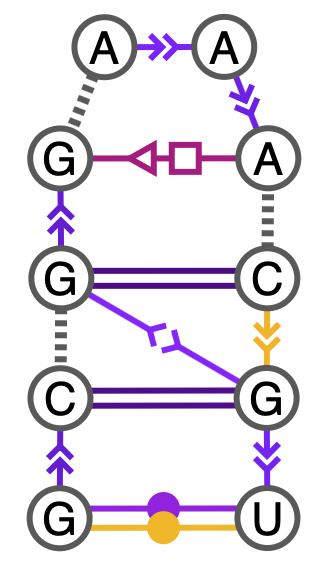}
	\caption{}
	    \end{subfigure}\vfill
	\begin{subfigure}[t]{0.33\textwidth}
	        \includegraphics[scale=0.20]{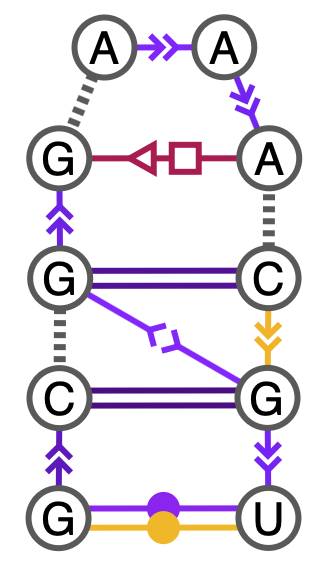}
	    \caption{}
	    \end{subfigure}\hfill
	\begin{subfigure}[t]{0.33\textwidth}
	        \includegraphics[scale=0.20]{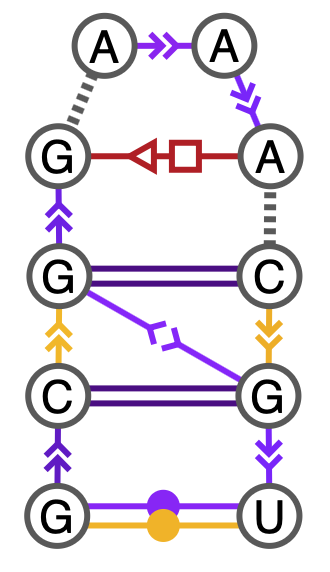}
	    \caption{}
	    \end{subfigure}\hfill
	    \begin{subfigure}[t]{0.33\textwidth}
		\includegraphics[scale=0.22]{dyn_colormap.png}
	    \end{subfigure}
		\caption{Dynamic secondary structures showing the probability of annotated interactions within grand-canonical reweighted ensembles
		at different values of reagent concentration: (a) $\SI{e-3}{\milli\Molar}$, (b) $\SI{0.017}{\milli\Molar}$,
		(c) $\SI{0.4}{\milli\Molar}$, (d) $\SI{2.7}{\milli\Molar}$, (e) $\SI{3.9}{\milli\Molar}$, (f) $\SI{5.7}{\milli\Molar}$,
		(g) $\SI{8.3}{\milli\Molar}$ and (h) $\SI{10}{\milli\Molar}$.
                A slight decrease in the population of the trans Hoogsteen/Sugar edge pair
                can be seen at higher concentration, consistent with the fact that the reagent can affect the tetraloop structure.
		}
	\label{dynstru_conc}
	\end{figure*}
\clearpage
	\section{Effect of grand-canonical reweighting}
Here we compare results obtained estimating the reactivities from simulations at constant number of particles
with reactivities obtained from the reweighting of the concatenated trajectories according to the grand-canonical ensemble.
Figure~\ref{sec:grand-canonical} shows that results are qualitatively comparable in a range where the number of
reagent copies has been chosen consistently with the enforced reagent concentration.
However, an important advantage of the grand-canonical reweighting procedure is that the behavior as a function
of concentration is much smoother than the behavior as a function of the number of reagent copies.
This is expected, since individual trajectories are subject to statistical errors. The possibility to
compute weighted averages from a single set of trajectories thus makes the estimation of concentration-dependent quantities
more statistically robust.
\label{sec:grand-canonical}
	 \begin{figure*}
                \centering
                \includegraphics[width=1.05\linewidth, keepaspectratio]{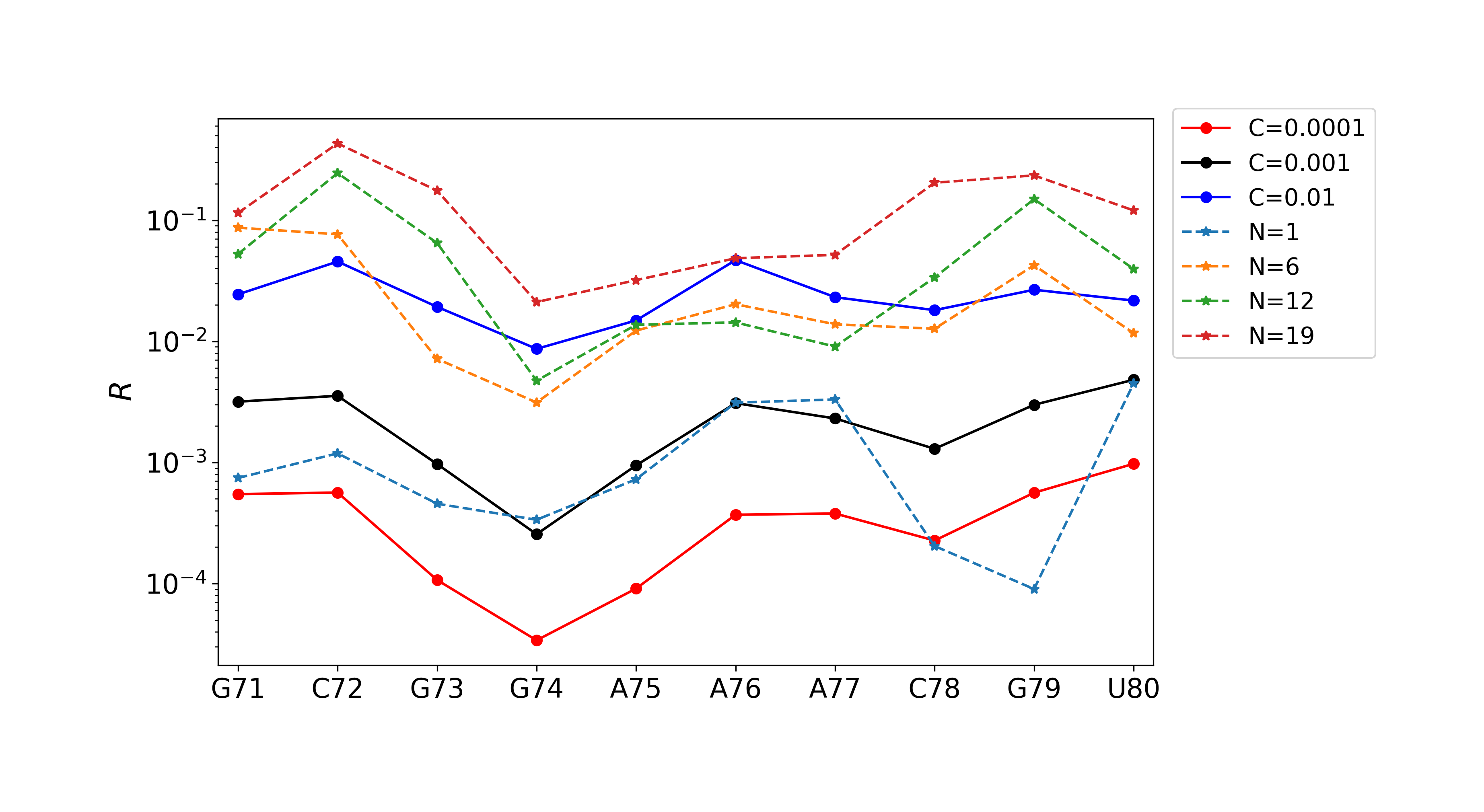}
                \caption{Reactivity profiles from simulations at fixed number of copies $N=1$, $6$, $12$, $19$ and at 
		 fixed concentrations $C = \SI{e-1}{\milli\Molar}$, $\SI{1}{\milli\Molar}$,
                 $\SI{10}{\milli\Molar}$ as obtained through grand-canonical reweighting.
		 }
        \end{figure*}
\clearpage
	\section{Effect of ionic conditions}
\label{sec:ionic-conditions}
As a control of the possible effects of different ionic conditions 
on the cooperative binding dynamics of RNA with 1M7, we generated a
trajectory at number of reagent copies fixed to a representative value of $N=6$,
in which the simulation environment contained additional 26 Cl$^-$/Na$^+$ ion pairs,
corresponding to a nominal concentration of $\SI{0.10}{\Molar}$,
in such a way that the total charge of the system was preserved. As
shown in Fig. \ref{counterions}, changes in the reactivity profile obtained
from the control simulation were of the order of differences associated 
with small ($\pm 1$) variations of the number of reagent copies.
This result is expected since SHAPE reagents are neutral, hence their distribution
around RNA should not be highly affected by the effective electrostatic screening.
	 \begin{figure*}
         	\centering
                \includegraphics[width=1.05\linewidth, keepaspectratio]{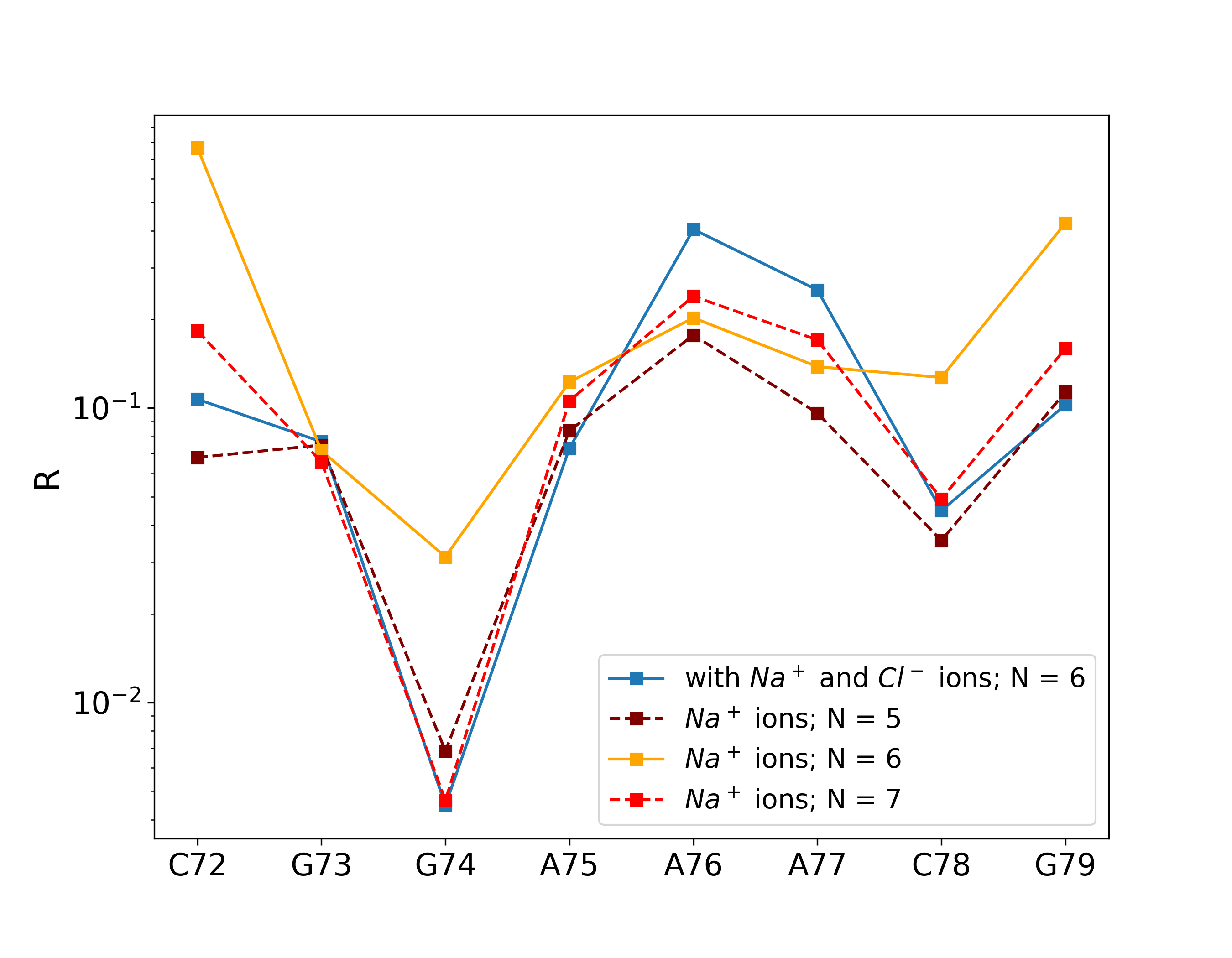}
		 \caption{Comparison of reactivity profiles at fixed number of reagent copies $N=6$ 
		  between two different ionic conditions: with only $Na^+$ ions
		 (blue) and with ions $Na^+$ and counter-ions $CL^-$ (orange). As further control,
		 reactivities in presence of only $Na^+$ ions with $N=5$ (maroon) and 
		 $N=7$ (red) reagent copies are reported.}
	\label{counterions}
        \end{figure*}
\clearpage
	\section{Effect of helix length}
\label{sec:longer-helix}
As a control on the possible bias introduced by the specific choice of
the length of the simulated helix, we generated a trajectory with an extended
portion of the SAM-I riboswitch gcgGAAAcgu tetraloop, namely ranging from
C69 to G82 (two additional base-pairs C69-G82 and A70-U81), fixing the number
of reagent copies to the representative value $N=6$. We thus compared
the reactivity profile computed from this control simulation with those reported
in our main study. As shown in Fig. \ref{longerhelix}, the reactivity profile
of the longer helix was consistent with the fluctuactions associated with the
varying number of reagent copies when simulating the shorter helix, as well as the 
average reactivity of nucleotides in the loop.
	\begin{figure*}\
                \centering
                \includegraphics[width=1.05\linewidth, keepaspectratio]{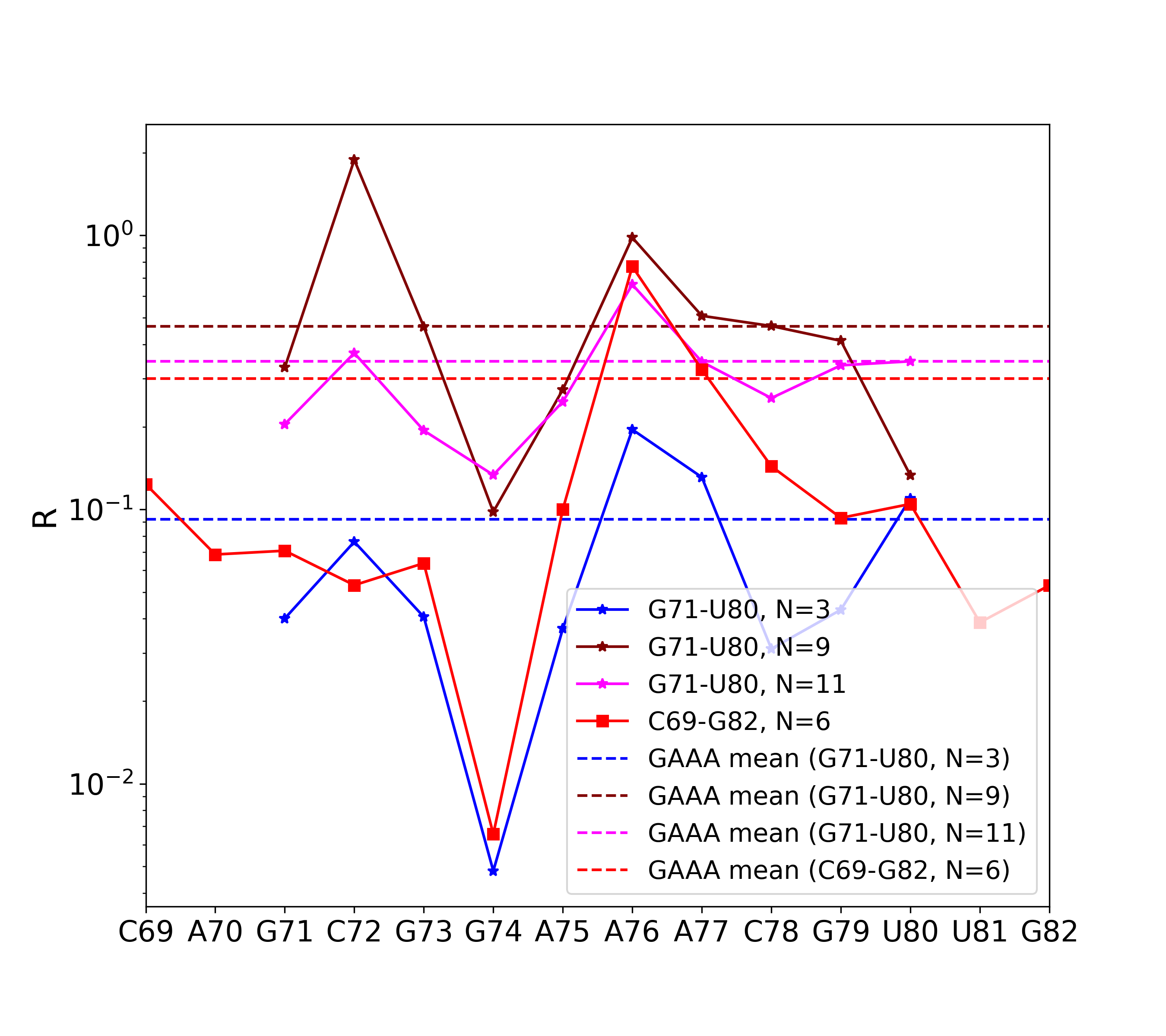}	
                \caption{Comparison of reactivity profiles at fixed number of reagent copies $N=6$ 
		between a longer C69-G82 helix and the shorter G71-U80 helix used in the main study.}
	\label{longerhelix}
        \end{figure*}
\end{suppinfo}

\clearpage
\bibliography{biblio}

\end{document}